\documentclass{aa}  

\usepackage[super]{nth}
\usepackage{graphicx}
\usepackage{placeins}
\usepackage{multirow}
\usepackage[separate-uncertainty=true, retain-unity-mantissa = false]{siunitx}
\usepackage{makecell}
\usepackage{tabularx}
\usepackage{overpic}
\usepackage[most]{tcolorbox}

\DeclareSIUnit\angstrom{\text {Å}}
\DeclareUnicodeCharacter{0301}{\'{e}}

\newcommand{\figref}[1]{\figurename~\ref{#1}}
\newcommand{\secref}[1]{Sec.~\ref{#1}}
\newcommand{\tabref}[1]{\tablename~\ref{#1}}

\newcommand{\ptre}{$\mathrm{PAH_{3.3}}\ $}
\newcommand{\<}{\hspace{0.3ex}}
\newcommand{\dd}{\mathrm{d}}

\usepackage[varg]{txfonts}

\usepackage[dvipsnames]{xcolor}
\definecolor{blue_dark}{rgb}{0.2, 0.2, 0.6}
\definecolor{blue_green}{rgb}{0.1, 0.4, 0.3}
\usepackage[hidelinks,colorlinks=true,linkcolor={blue_dark},citecolor={blue_dark}, urlcolor={blue_dark},filecolor={blue_dark}]{hyperref}
\usepackage{orcidlink}

\newcounter{subfig}[figure]

\newcommand{\subfiglabel}[1]{\refstepcounter{subfig}\label{#1}}

\begin{document} 
   \title{PAH suppression in galaxy cluster environment at $z\sim0.3$}

   \author{Pietro Benotto\orcidlink{0009-0004-7602-0160}\inst{1,2}\fnmsep\corrauth{pietro.benotto@inaf.it}
   \and
   Benedetta Vulcani\orcidlink{0000-0003-0980-1499}\inst{1}
   \and
   Giulia Rodighiero\orcidlink{0000-0002-9415-2296}\inst{2,1}
   \and
   Peter J. Watson\orcidlink{0000-0003-3108-0624}\inst{1,3}
   }

   \institute{
        \inst{1} INAF, Osservatorio Astronomico di Padova, Vicolo dell’Osservatorio 5, 35122, Padova, Italy.\\
        \inst{2} Dipartimento di Fisica e Astronomia "Galileo Galilei", Università di Padova, Vicolo dell'Osservatorio 3, 35122 Padova, Italy. \\
        \inst{3} Universit\'e C\^ote d'Azur, Observatoire de la Côte d’Azur, CNRS, Laboratoire Lagrange, 06000 Nice, France
    }
   
   \date{}
    
   \abstract{Polycyclic aromatic hydrocarbons (PAHs) are widespread molecules made up of aromatic rings, whose emission bands dominate the mid-infrared spectra of star-forming galaxies. In extragalactic astronomy, PAH emission is often used to trace obscured star formation, although the reliability of this tracer depends on galaxy properties. In this work, we perform a statistical analysis of how the cluster environment affects PAH 3.3\,$\mathrm{\mu m}$ emission, focusing on redshift $0.195< z< 0.415$. In order to achieve this, we use a large sample of HST and JWST/NIRCam imaging, tracing PAH emission through JWST medium-band filters and deriving galaxy properties using multiband SED fitting. We analyse the clusters Abell~2744, A370, and MACS0416 and compare the results with the COSMOS, GOODS-S, and UDS fields. We uncover that just the $(6\pm1)\%$ of cluster galaxies in our sample are PAH emitters compared to the $(42\pm4)\%$ of field galaxies. The result follows the different ratio of star-forming and quiescent galaxies in the two environments. However, we also find that galaxies in the UVJ transition zone, between the star-forming and quiescent regions, show the strongest environmental effects: in this region, the fraction of PAH emitters with mass below $10^{10.2}\,\mathrm{M_\odot}$ is $1_{-1}^{+3}\%$ in clusters, compared to $63_{-12}^{+10}\%$ in the field, showing a $5\sigma$ significance contrast. Given that we find no evidence for a difference in the star-formation histories, this difference may be due to a direct effect of the environment on the galaxies' interstellar medium (ISM). This hypothesis is corroborated by the results of the analysis of the dust attenuation in these galaxies. These results demonstrate that calibrations of PAH emission as an SFR tracer must account for environment, not just intrinsic galaxy properties, and highlight how the cluster environment can directly shape the PAH component of the ISM faster than star quenching.}
   
   \authorrunning{P. Benotto et al.}
   \maketitle
\nolinenumbers
\section{Introduction}
\label{sec:introduction}
 
Polycyclic aromatic hydrocarbons (PAHs) are large, planar molecules built from aromatic rings of carbon, saturated by hydrogen atoms \citep[e.g.][]{Tielens:2008, Li:2020}. Through a rich set of vibrational features spanning $\SIrange{3}{20}{\micro\meter}$, they dominate the mid-infrared spectra of an extraordinarily wide range of astrophysical environments, from galactic photodissociation regions \citep{Knight:2022, VanDePutte:2025} and the circumstellar envelopes of AGB stars \citep{Hony:2003, Matsuura:2014} to the diffuse interstellar medium (ISM) \citep{SLoan:1999, Murgia:2026} and entire galaxies \citep{Smith:2006, Rigopoulou:2024}, with PAHs being detected up to $z=4.2$ \citep{Spilker:2023}.

PAHs contain a substantial fraction of the interstellar carbon budget and account for roughly 5--20 per cent of the total infrared luminosity of a galaxy \citep{Tielens:2008, Tielens:2026}. Beyond their role as a dust component, PAHs are central to the thermal balance of the ISM, both shaping the structure of PDRs and regulating the conditions that lead to molecular cloud formation \citep{Bakes:1994, Hollenbach:1999, Weingartner:2001, Tielens:2008}. The PAH abundance and charge state are continually changed by injection from evolved-star winds and supernova ejecta, and reprocessed by shocks and radiation fields. For this reason, the spectral properties of PAHs are broadly linked to the life cycle of interstellar dust \citep{Allamandola:1985, Allamandola:1999, Tielens:2008}.
 
Among the PAH bands, the one at $\SI{3.3}{\micro\meter}$ was identified, through ground-based $L$-band spectroscopy, as the C--H stretching mode of small, predominantly neutral PAHs \citep{Tokunaga:1991, vanDiedenhoven:2004, Tielens:2008}. Low- and moderate-resolution observations with subsequent space infrared missions, including ISO, AKARI and Spitzer, extended the study of this feature to samples of nearby star-forming galaxies and active galactic nuclei \citep[AGN;][]{Moorwood:1986, Imanishi:2008, Murata:2014}. However, its intrinsic weakness relative to the longer-wavelength PAH bands, together with its proximity to a telluric water absorption feature that severely limits ground-based observation, left the $\SI{3.3}{\micro\meter}$ band comparatively unexplored for decades. The unprecedented mid-infrared sensitivity and spatial and spectral resolution of JWST \citep{Gardner:2006} have rectified this situation, making the $\SI{3.3}{\micro\meter}$ feature an accessible and well-studied PAH band \citep[e.g.][]{Sandstrom:2023, Chastenet:2023, Lai:2023, Boersma:2023}.
 
Because PAHs efficiently absorb ultraviolet (UV) photons and re-radiate the absorbed energy in the mid-infrared, they are widely used as tracers of star formation in dust-obscured environments \citep{Forster:2004, Peeters:2004, Shipley:2016, Xie:2019}. However, PAH  emission strength is not governed solely by the recent star-formation rate (SFR). It also critically depends on the size and ionisation state of the emitting molecules, as well as on the physical conditions of the surrounding medium, as energetic shocks and hard radiation fields can efficiently destroy the smallest PAHs \citep{Micelotta:2010, Slavin:2015, Geballe:1989, Herrero:2014, Rigopoulou:2024}.
Low metallicity environments also systematically suppress PAH emission \citep{Lai:2020, Whitcomb:2024}. 

\citet{Lyu:2025} carried out a statistical analysis of field galaxies at $0.2<z<0.5$ using JWST NIRCam Wide Field Slitless Spectroscopy, providing an unbiased sample with which to test the reliability of PAH emission as a star-formation indicator. They report a correlation between the fraction of PAH emitters and galaxy morphology. Taken together, these studies indicate that using PAH emission as a star-formation tracer robustly requires a detailed characterisation of its dependence on host-galaxy properties.
 
However, the environment surrounding galaxies can also leave a clear imprint on their PAH emission. \citet{Vulcani:2023, Vulcani:2025} identified a population of galaxies in and around the Abell 2744 cluster with an excess of flux in the F444W NIRCam filter. This excess was shown to be driven, at least in a subset of cases with NIRSpec observations, by strong $\SI{3.3}{\micro\meter}$ PAH emission. Working with NIRCam photometry of the same cluster, \citet{Cheng:2025} analysed a sample of $\SI{3.3}{\micro\meter}$ PAH emitters and concluded that most of the emitting galaxies are systems that have only recently fallen into the cluster and still host residual star formation, or have only recently quenched. The study, however, lacked a direct comparison with a coeval field population, which is needed to assess whether the inferred PAH properties are shaped by the cluster environment or simply reflect the properties of star-forming galaxies in general. Complementing this picture, \citet{Benotto:2026a} mapped the $\SI{3.3}{\micro\meter}$ emission of galaxies undergoing ram-pressure stripping \citep[RPS;][]{Gunn:1972} in Abell 2744, and detected PAH emission not only in the galactic disks but also in star-forming clumps embedded in the stripped tails, up to several tens of kpc away. This shows that PAH-carrying grains can be stripped together with the neutral gas and survive both the heating of the intracluster medium (ICM) and the shocks generated as the stripped ISM interacts with it.
 
Despite this growing evidence for environmental effects on PAH molecules in clusters, no study has yet systematically characterised the global statistical properties of PAH emission across a cluster galaxy population, homogeneously comparing them with a field sample. Closing this gap is the main goal of the present work: by relating PAH strength to galaxy properties in both cluster and field environments, we assess whether cluster membership has a direct effect on a galaxy's global PAH emission, comparing cluster galaxies with field galaxies matched in colour, stellar mass and star-formation rate in a statistical sense.
 
The paper is organised as follows. In Sect.~\ref{sec:methods} we present the data set and the methods adopted to select and characterise the PAH emission; Sect.~\ref{sec:results} presents our results; Sect.~\ref{sec:discussion} discusses their implications; and Sect.~\ref{sec:conclusions} summarises our conclusions. Throughout this work we adopt a \citet{Kroupa:2001} initial mass function and a standard $\Lambda$CDM cosmology with $\Omega_\mathrm{m}=0.3$, $\Omega_\mathrm{\Lambda}=0.7$, and $H_0=\SI{70}{km\,s^{-1}\,Mpc^{-1}}$.

\section{Methods}\label{sec:methods}
\subsection{Available data}\label{sec:selection}
\begin{figure*}
    \centering
    \includegraphics[width=0.49\textwidth]{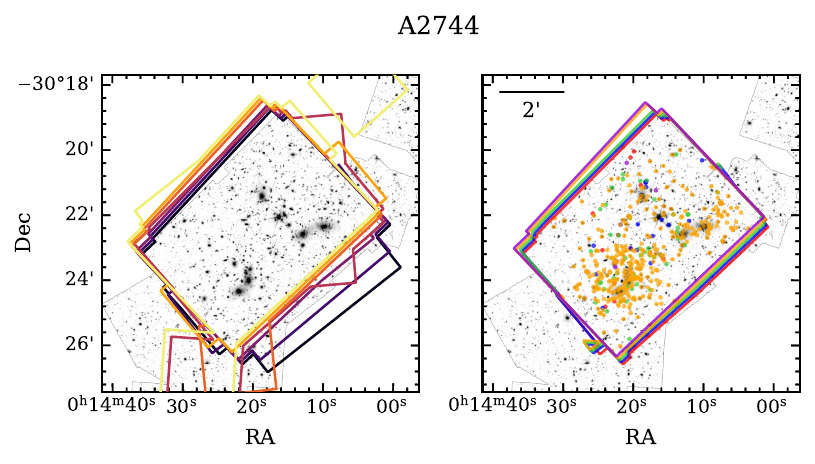}
    \includegraphics[width=0.49\textwidth]{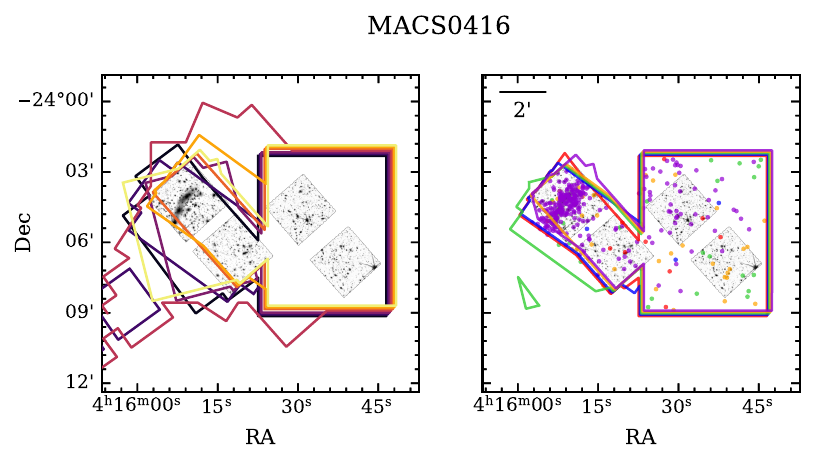}
    \includegraphics[width=0.49\textwidth]{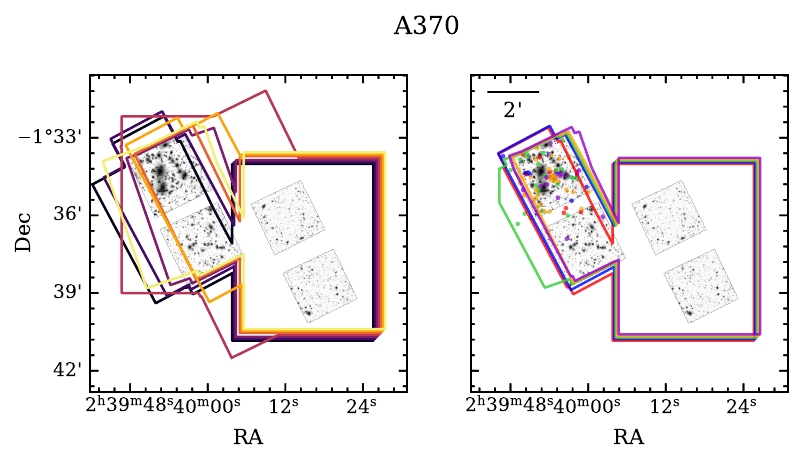}
    \includegraphics[width=0.49\textwidth]{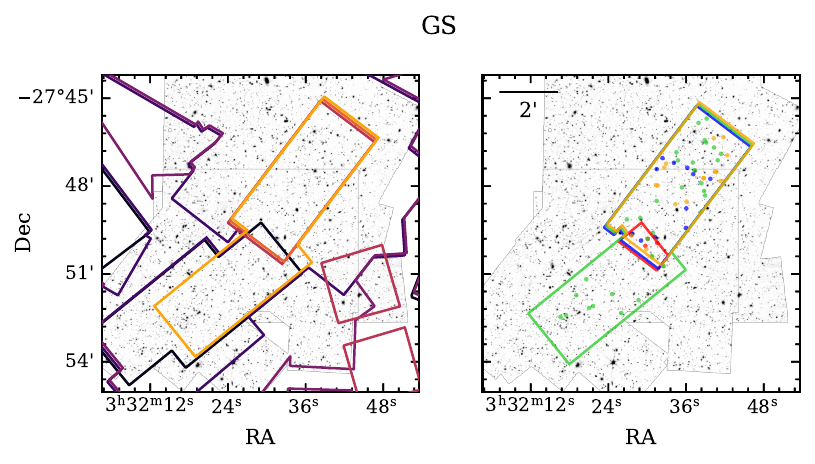}
    \includegraphics[width=0.49\textwidth]{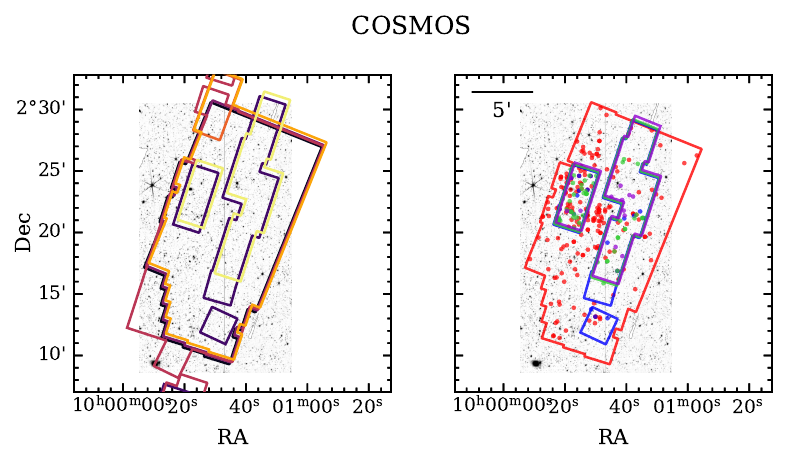}
    \includegraphics[width=0.49\textwidth]{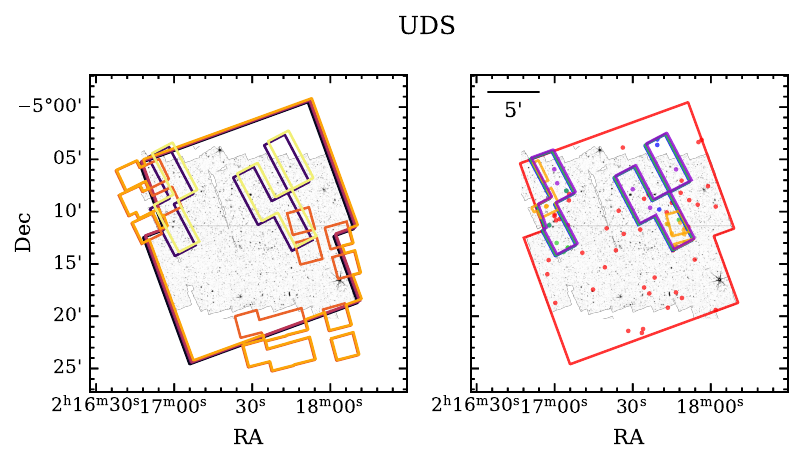}\\\vspace{-2mm}
    \hspace*{6mm}\includegraphics[width=0.47\textwidth]{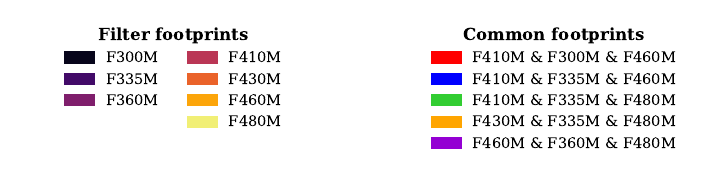}
    \hspace{4mm}\includegraphics[width=0.47\textwidth]{img/map_A2744_legend.pdf}
\caption{Available images and galaxies in each analysed field. In each one, the left column displays the MAST footprints for each filter used to extract the PAH emission in the redshift range $0.195 < z < 0.415$. The right column shows the regions where combinations of filters are used to detect PAH at different redshifts. Each colour represents areas where all three filters are contemporaneously available. Circles of the same colour indicate the galaxies that are at the appropriate redshift and within the regions covered by the filter combinations, according to the footprints. The F444W JWST image is shown in the background.}
\label{fig:available-filters}
\end{figure*}
We use JWST/NIRCam medium-band photometry to quantify PAH emission in galaxies within the spectroscopic redshift interval $0.195<z<0.415$\footnote{Our probed redshift range is set by the redshift at which PAH emission enters the F410M filter (lower bound) and exits the F460M filter (upper bound), as discussed in \secref{sec:pah-extraction}.}. Given that PAH flux extraction requires observations in multiple medium-band filters, as described in \secref{sec:pah-extraction}, and that many more filters are necessary for accurate galaxy SED fitting, we exploit extensively studied sky regions with abundant HST and JWST archival data.

For each region, we identify a primary online source for the available ready-to-use mosaics and then recompute the mosaics ourselves for the medium-band filters used to extract the \ptre emission (F300M, F335M, F360M, F410M, F430M, and F480M; see \secref{sec:pah-extraction}). This recomputation incorporates all recent observations (up to July 2026) and is performed using \href{https://github.com/gbrammer/grizli}{\textsc{grizli}} \citep{10.5281/zenodo.1146904}, maximising the sample of galaxies for which \ptre emission can be measured. Conversely, recomputing the mosaics for all bands to add a few bands to the SED fitting would give marginal benefits, given the already available large number of photometric points for every galaxy. Following this approach, we selected the A2744, MACS0416, A370, COSMOS, UDS, and GOODS-S sky regions, which provide the filter combinations required for our PAH analysis\footnote{Although MACS1149 and GOODS-N also provide the necessary NIRCam medium-band photometry to extract \ptre, we exclude these fields due to the small number of spectroscopically confirmed galaxies in our chosen redshift range ($N_\mathrm{spec} < 20$).}. 
Table~\ref{tab:fields} summarises the selected fields, their coordinates, and the corresponding galaxy counts.

\begin{table*}[htb]
    \caption{\centering Fields used in this work and number of selected galaxies per field.}
    \label{tab:fields}
    \centering
    \begin{tabular}{lcccccc}
    \hline\hline
    Field name & Cluster/field & RA [hms] & DEC [dms] & $z_\mathrm{cl}$ & $N_\mathrm{spec}$ & $N_\mathrm{good}$\\
    \hline
    A2744 & cluster & \multirow{2}*{00:14:20.2} & \multirow{2}*{-30:22:26} & [0.284, 0.325] & 365 & 259 \\
    A2744 & field &  &  & - & 82 & 18 \\
    A370 & cluster & \multirow{2}*{02:39:54.8} & \multirow{2}*{-01:35:45} &  [0.350, 0.400] & 8 & 2\\
    A370 & field &  &  & - & 78 & 17 \\
    MACS0416 & cluster & \multirow{2}*{04:16:13.0} & \multirow{2}*{-24:04:59} &  [0.382, 0.412] & 293 & 164\\
    MACS0416 & field &  &  & - & 74 & 17 \\
    COSMOS & field & 10:00:32.6 & +02:21:18 & - & 276 & 102 \\
    GS & field & 03:32:27.4 & -27:49:37 & - & 68 & 3 \\
    UDS & field & 02:17:28.3 & -05:08:38 & - & 75 & 24 \\
    \hline
    \end{tabular}
    \tablefoot{$z_\mathrm{cl}$ indicates the redshift range used to identify a galaxy as a cluster member, $N_\mathrm{spec}$ is the total number of spectroscopic objects for which it is possible to extract the \ptre map, and $N_\mathrm{good}$ is the selected number of objects after manually removing AGNs or unsuitable sources (see \secref{sec:sources}) and after the mass-cut, as described in \secref{sec:completeness}.}
\end{table*}

A2744 is part of the Hubble Frontier Fields \citep{Lotz:2017}, with observations including HST WFC3/UVIS, WFC3/IR, and ACS/WFC, alongside extensive JWST photometry \citep{Treu:2022, Paris:2023, Merlin:2024, Bezanson:2024, Weaver:2024, Suess:2024}. This is the only field where we do not recompute the red medium-band filter mosaics, as the releases from the UNCOVER and Megascience surveys \citep{Bezanson:2024, Suess:2024} are fully up to date; we only supplement them with the F335W mosaic from the \href{https://dawn-cph.github.io/dja/imaging/v7/}{DAWN JWST Archive} \citep[DJA,][]{Valentin0:2023}.
MACS0416, another Hubble Frontier Field \citep{Lotz:2017}, features multi-filter HST coverage and JWST wide/medium-band data from the CANUCS and Tecnicolor surveys \citep{Sarrouh:2025}. We further incorporated observations from programs 1176 \citep[PI: Windhorst,][]{Windhorst:2023}, 2883 (PI: Sun), 3538 (PI: Iani), and 5890 (PI: Withers) covering the central cluster region in the redder JWST medium bands. A370 is similarly covered by the CANUCS and Tecnicolor surveys \citep{Sarrouh:2025}, with central cluster observations in the red medium filters provided by programs 2883 (PI: Sun), 3538 (PI: Iani), and 5890 (PI: Withers).

For COSMOS, UDS, and GOODS-S, primary mosaics are sourced from the DJA and updated with all available redder medium-band JWST exposures. In COSMOS, we include observations from programs 1837 (PI: Dunlop), 6434 (PI: Egami), 6368 (PI: Dickinson), and 7814 (PI: Muzzin). For UDS, we incorporate data from programs 1837 (PI: Dunlop), 3993 (PI: Morishita), 6368 (PI: Dickinson), and 7814 (PI: Muzzin). In GOODS-S, although observations are available via JADES DR4 \citep{Eisenstein:2026, Johnson:2026}, we utilise the DJA mosaics to ensure consistent data reduction across fields and to retain legacy HST observations not included in JADES.

Together, these fields offer a total area of approximately 550 square arcminutes with continuous photometric coverage spanning from \SI{200}{nm} to nearly $\SI{5}{\micro m}$. Every galaxy in our sample is detected in at least 15 distinct filters, with over half detected in more than 22. This comprehensive wavelength coverage allows us to trace galaxy SEDs with exceptional precision. The complete list of filters used is provided in Appendix~\ref{app:filters}.

\label{sec:sources}
Together with the imaging data, we compile spectroscopic redshifts from multiple surveys to identify galaxies for which PAH emission can be measured.
The spectroscopic catalogues used in each field are listed in \tabref{tab:specz_cats}. Where quality flags are provided, we retain only secure or likely redshifts, following each survey's confidence thresholds.
For each location in each field, we examine the available filters and retain only the galaxies with a redshift that aligns with the range compatible for extracting the \ptre emission, as outlined in \secref{sec:pah-extraction}.
For the A2744, A370 and MACS0416 fields, we consider a galaxy to be a member of the cluster if its redshift is in the range $0.284\le z_{cl}<0.325$ for A2744 \citep{Vulcani:2023}, $0.350\le z_{cl}<0.400$ for A370 \citep{Lagattuta:2022}, and $0.382\le z_{cl}<0.412$ for MACS0416 \citep{Ragusa:2025}. The total number of galaxies available in each field is displayed in the column $N_\mathrm{spec}$ of \tabref{tab:fields}.

We also perform a visual inspection of each source to exclude unsuitable galaxies, e.g. galaxies cut because they are near the sensor border, galaxies heavily contaminated by other galaxies or from bright stellar diffraction spikes. We are primarily interested in the SFR--PAH relation, so by comparing our selected galaxies with the literature, SIMBAD, and NED, we also remove any known AGN-contaminated galaxies in our sample (22 galaxies in total). Specifically, we exclude sources with confirmed X-ray detections \citep[from][]{2011ApJS..195...10X, 2012MNRAS.427.3103B, 2012A&A...542A..16R, 2016ApJS..224...40W}, objects showing signatures of AGN from optical, infrared, or spectral analysis or variability \citep[from][]{2012MNRAS.427.3103B, 2012MNRAS.426..360V, 2017ApJS..228....2L,  Moretti+22, Watson:2025}, and galaxies identified through their radio emission \citep[from][]{2013wise.rept....1C, 2016MNRAS.455.2731R, 2017A&A...602A...1S, 2018MNRAS.478.3848M, 2019A&A...627A.142V, 2020ApJ...903..139A}, alongside additional radio galaxies identified directly via SIMBAD.

\begin{table*}[htb]
\centering
\caption{\centering Spectroscopic catalogues used in each field.}
\label{tab:specz_cats}
\begin{tabular}{lp{6cm}p{3cm}}
\hline\hline
Catalogue & Reference & Fields \\
\hline\\[-0.8em]
ASTRODEEP          & \citet{Merlin:2024}                             & A2744, COSMOS, UDS, GOODS-S \\
All the Little Things & \citet{Naidu:2024}                           & A2744 \\
UNCOVER            & \citet{Prince:2024}                             & A2744 \\
GLASS-JWST NIRISS  & \citet{Watson:2025b}                            & A2744 \\
MUSE (archival)    & \citet{Owers:2011, Braglia:2009, Bergamini:2023}& A2744 \\
MUSE A2744         & \citet{Torralba:2026}                           & A2744 \\
GLASS              & \citet{Schmidt:2014, Treu:2015}                 & A2744, MACS0416, A370 \\
CANUCS             & \citet{Sarrouh:2025}                            & MACS0416, A370 \\
MUSE clusters      & \citet{Richard:2021}                            & MACS0416, A370 \\
VIMOS              & \citet{Caminha:2017}                            & MACS0416 \\
HST/MUSE A370      & \citet{Lagattuta:2022}                          & A370 \\
COSMOS compilation & \citet{Khostovan:2025}                          & COSMOS \\
DESI-XMM           & \citet{Ratajczak:2026}                          & UDS \\
X-ray to FIR       & \citet{Zou:2022}                                & UDS \\
HLF                & \citet{Whitaker+2019}                           & GOODS-S \\
MUSE HUDF          & \citet{Bacon+2023}                              & GOODS-S \\
MuseWide           & \citet{Urrutia:2019}                            & GOODS-S \\
3D-HST             & \citet{Momcheva:2016}                            & All fields \\
DJA                & \citet{DJA_zenodo}                              & All fields \\
\hline
\end{tabular}
\end{table*}

\subsection{PSF matching, background modelization, galaxy disk}\label{sec:pah-extraction}
We first scale down all the mosaics to a common resolution of $\SI{0.04}{\arcsec/px}$, which corresponds to $\SI{130}{pc}$--$\SI{220}{pc}$ per pixel in the $0.195<z<0.415$ redshift range. We then build the ePSF (effective point spread function) for our mosaics using \href{https://github.com/pietrobenotto/EPIMETHEUS}{\textsc{epimetheus}} \citep{Benotto:2026b}. We combine all the mosaics to compute an average ePSF, and then, for each mosaic, we use the corresponding ePSF aligned with the most frequent exposure orientation in that mosaic. We do not use the ePSFs released with some of the data used --- such as UNCOVER, CANUCS, JADES and COSMOS --- as not every filter is covered, and the procedure used to produce them is not uniform across different fields.

Subsequently, we matched the PSF of all HST and JWST mosaics to that of the F480M filter, using the \href{https://pypher.readthedocs.io/en/latest/}{\textsc{PyPHER}} tool with a regularisation parameter of $3\cdot10^{-3}$ on $3\times$ oversampled ePSFs, after which the kernels are reduced to the original pixel scale, following the procedure used in \citet{Weaver:2024} and \citet{Benotto:2026b}. The resulting PSF full-width at half maximum (FWHM) is $\SI{4.4}{px}$ ($\SI{0.18}{arcsec}$). We also correct the mosaics for galactic dust extinction along the line of sight, with $E(B-V)$ values obtained from the \citet{Chiang:2023} dust map and adopting a \citet{Fitzpatrick:1999} extinction law with $R_V=3.1$. We model and subtract the background in each filter and for each galaxy by creating a $20\arcsec \times 20\arcsec$ cutout centred on the galaxy and using the \href{https://photutils.readthedocs.io/}{\textsc{photutils}} Background2D class.

To ensure consistent photometry across all bands, we define a single spatial region of interest (ROI) for each galaxy based on the F200W image. The ROI is determined by selecting the contiguous pixels with flux levels above $5\sigma$ relative to the background noise, and by fitting an ellipse to this region. Rather than adopting the segmentation map directly, we opt for this approach to ensure that the source is not contaminated by foreground or background objects, which could otherwise lead to false-positive PAH detections. To this end, we perform a visual inspection of each galaxy and manually mask any regions affected by light from adjacent sources. The resulting masked ROI is then applied consistently to extract total fluxes across all photometric bands, and the total PAH flux associated with each galaxy is defined as the emission falling within this region. This study investigates in-disk PAH emission to understand the mechanisms affecting galactic disks. While \citet{Benotto:2026a} noted that some galaxies with RPS may show extraplanar PAH emission, we do not consider that emission here in the total PAH galactic flux.

\subsection{Fields completeness}\label{sec:completeness}
We calculate the spectroscopic completeness by estimating the fraction of galaxies with spectroscopic redshifts that fall within the redshift range of interest, relative to the total number of galaxies in the same redshift range, as a function of magnitude.

We start from the number of galaxies with photometric redshifts in the range ($N_\mathrm{ph}$). This number can be expressed as
\begin{align}
    N_\mathrm{ph} = q\, N_\mathrm{tr} + f\, N_{\overline{\mathrm{tr}}}
\end{align}
where $N_\mathrm{tr}$ is the true, unknown number of galaxies in the range of interest and $N_{\overline{\mathrm{tr}}}$ is the true number of galaxies outside the same range. $q$ and $f$ are the detection efficiency and false positive fraction, respectively. Given also $k_{\mathrm{cl}} = N_\mathrm{ph} / N$, where $N = N_\mathrm{tr} + N_{\overline{\mathrm{tr}}}$ is the total number of galaxies, $N_\mathrm{tr}$ can be estimated as:
\begin{align}
    N_\mathrm{tr} = \frac{k_\mathrm{cl} - f}{q - f}\, N_\mathrm{ph}
\end{align}
For each field and each magnitude range (1 magnitude wide), we evaluate $q$ and $f$ using the spectroscopic subsample as:
\begin{align}
    q &= \frac{N_{\mathrm{ph,sp}}}{N_\mathrm{sp}} \\
    f &= \frac{N_{\mathrm{ph,\overline{sp}}}}{N_{\overline{\mathrm{sp}}}}
\end{align}
where $N_\mathrm{sp}$ and $N_{\overline{\mathrm{sp}}}$ are the number of spectroscopic galaxies inside and outside the redshift range, respectively; $N_{\mathrm{ph,sp}}$ is the number of galaxies in range according to both photometric and spectroscopic redshift; and $N_{\mathrm{ph,\overline{sp}}}$ is the number of galaxies in range by photometry but outside by spectroscopy.

As photometric redshift catalogues, we use ASTRODEEP \citep{Merlin:2024} for A2744, COSMOS, UDS, and GOODS-S, and CANUCS \citep{Sarrouh:2025} for A370 and MACS0416, adopting the F200W Kron magnitude of each object for the completeness measurements as a function of magnitude. We rely on these publicly available catalogues rather than recomputing photometric redshifts from our own photometry, since, although individual objects may differ, the resulting redshift distributions are expected to be statistically consistent with the published ones. This is sufficient for our purposes, as these catalogues are only used to estimate the completeness of our spectroscopically confirmed sample.
\begin{figure}
    \centering
    \includegraphics[width=0.7\linewidth]{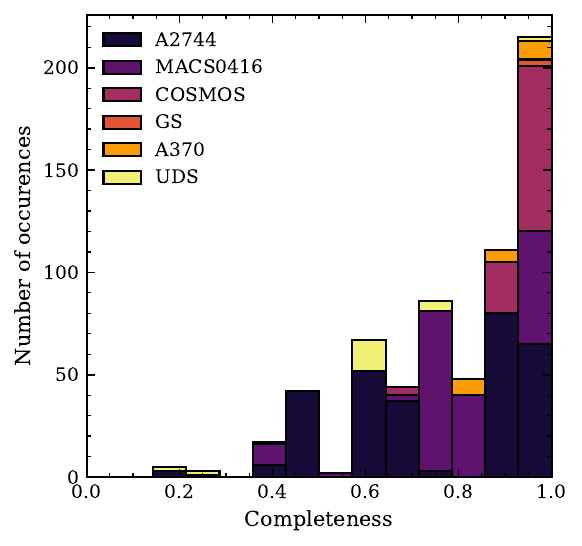}
    \caption{Stacked histogram of the completeness distribution for each field for every galaxy with mass higher than $M_\mathrm{compl}$}
    \label{fig:completeness}
\end{figure}
In the magnitude bin between magnitudes 22 and 23, the completeness of the clusters is $\sim70\%$ for MACS0416, $100\%$ for A370 and $\sim50\%$ for A2744. At the same magnitude, UDS is $\sim15\%$ complete, while COSMOS and GOODS-S are respectively $\sim80\%$ and $100\%$ complete. As the completeness starts to drop sharply at higher magnitudes, we therefore select the magnitude limit of our sample to be $m_{\mathrm{lim}}  = 23$. The distribution of weights for the galaxies brighter than $m_{\mathrm{lim}}$  is illustrated in \figref{fig:completeness}.

To derive the stellar mass completeness limit of our sample, we consider that at any given stellar mass, quiescent galaxies exhibit the highest mass-to-light ratios due to their evolved stellar populations. Consequently, a flux-limited survey will lose completeness for quiescent systems before it does for actively star-forming galaxies. We estimate the mass limit of our sample using the masses and rest-frame $U, V, J$ colours derived from our SED fitting (see \secref{sec:sed}). We first select the quiescent population using the UVJ diagram \citep{Williams:2009, Whitaker:2011}. We then isolate the quiescent galaxies within the magnitude bin $[m_{\mathrm{lim}}, m_{\mathrm{lim}} + 1]$.Within this magnitude bin, we define our stellar mass completeness limit ($M_\mathrm{compl}$) as the $\nth{90}$ percentile of the quiescent galaxy mass distribution, which is $M_\mathrm{compl} = 10^{8.9} M_\odot$. We use $M_\mathrm{compl}$ as the mass lower limit of the galaxies analysed in this work.

\subsection{PAH flux extraction}
\begin{figure*}
    \includegraphics[width=\textwidth]{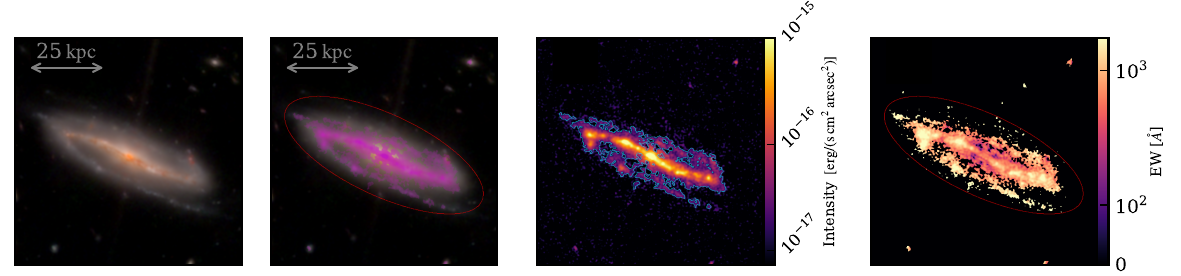}
    \caption{Example of PAH map for one galaxy in our sample. The first panel on the left shows the RGB composite image of the galaxy (F090W, F150W, F200W bands), the second panel shows the same RGB image with the extracted \ptre emission superimposed in magenta. The red ellipse represents the galaxy ROI. The third panel shows the \ptre intensity map. The blue contours indicate the region with SNR>5 and $\mathrm{EW}>\SI{200}{\angstrom}$ inside the ROI. The rightmost panel shows the EW of the \ptre emission and the galaxy ROI.}
    \label{fig:pah-extraction}
\end{figure*}
For each galaxy, given its redshift, we identify the NIRCam medium filter $\mathrm{F_{C}}$ in which the PAH falls, and two more filters, one on the left $\mathrm{F_L}$ and one on the right $\mathrm{F_R}$, to estimate the continuum contribution in the $\mathrm{F_{PAH}}$ filter.
\begin{table}
    \centering
    \caption{\centering Filters used to measure the \ptre flux}
    \label{tab:filters_pah}
    \begin{tabular}{c|ccc}
    \hline\hline
    Redshift range & $\mathrm{F_{C}}$ & $\mathrm{F_{L}}$ & $\mathrm{F_{R}}$\\
    \hline\\[-0.87em]
    $[0.195,0.225)$ & F410M & F300M & F460M \\
    $[0.225,0.240)$ & F410M & F335M & F460M \\
    $[0.240,0.285)$ & F410M & F335M & F480M \\
    $[0.285,0.335)$ & F430M & F335M & F480M \\
    $[0.390,0.415)$ & F460M & F360M & F480M \\
    \end{tabular}
\end{table}
Table \ref{tab:filters_pah} displays the filter combinations used as a function of the redshift. The lower and upper redshift limits for which to use a specific filter combination are defined as the redshifts where the peak of the \ptre emission coincides with the points on the filter transmission curve that correspond to half of its maximum transmission. The chosen filters to measure the continuum are a compromise between, in order of decreasing importance, filters availability, exclusion of the strong $\mathrm{Br\<\alpha}$ emission at $\SI{4.05}{\micro\meter}$, exclusion of the $\mathrm{H_2O}$ absorption band at $\SI{3.05}{\micro\meter}$ \citep{Lai:2020}, exclusion of the PAH plateau emission at $\SI{3.5}{\micro\meter}$ \citep{Lai:2020, Vulcani:2025}, and baseline minimisation.

The PAH flux is estimated from the flux densities measured in the three filters, $F_{\mathrm{L}}$, $F_{\mathrm{C}}$, and $F_{\mathrm{R}}$, where the two lateral filters are assumed to be dominated by the continuum. The continuum contribution within the central filter is estimated by linearly interpolating the flux densities measured in the two flanking filters. The PAH flux is then calculated as
\begin{align}
F_{\mathrm{PAH}} =
\left[f_{\mathrm{F_C}}-(1-k)f_{\mathrm{F_L}}-k f_{\mathrm{F_R}}\right]
\frac{c\int T_\lambda/\lambda\,\mathrm{d}\lambda}
{\lambda_{\mathrm{PAH}}T_{\lambda_{\mathrm{PAH}}}},
\end{align}
where $T_\lambda$ is the transmission curve of the central filter and $k$ is defined by the relative pivot wavelengths of the three filters. The full derivation is provided in Appendix~\ref{app:formula}.

We then obtain the $F_\mathrm{PAH}$ map together with its associated uncertainty, obtained by propagating the variance map of the filters used to obtain the PAH map. The pixels we consider to contain reliable PAH fluxes are those with a signal-to-noise ratio (SNR) greater than $5$ and with a PAH equivalent width (EW) greater than $\SI{200}{\angstrom}$. Moreover, to reduce the impact of bad pixels and stochastic noise, we only consider the \ptre emission to be reliable in regions of contiguous pixels larger than 15\footnote{15 pixels is the area of a circle with a diameter equal to the  PSF-convolved FWHM.}. An example of this extraction is shown in \figref{fig:pah-extraction}, where the combination of SNR and EW is visible. Given the bandpass width of the $F_\mathrm C$ filters used to detect PAH emission, the EW cut corresponds to a PAH contribution to the filter flux of at least $5\%$ to $10\%$ depending on the filter. For galaxies where no PAH emission is detected, we estimate an upper limit on its flux by determining the mean noise level within the galaxy ROI, and then calculating the total flux of a 15-pixel region whose signal would be $5\sigma$ above this mean background. This calculated flux is then adopted as the upper limit for that galaxy.

\subsection{SED Fitting}\label{sec:sed}
We used \href{https://bagpipes.readthedocs.io/en/latest/}{\textsc{bagpipes}} \citep{Carnall:2018} to perform the SED fitting for each galaxy by summing up the flux inside each galaxy's ROI for each PSF-matched and background-subtracted filter. To account for the systematic uncertainties in the fluxes \citep{Watson:2025}, we added a $5\%$ uncertainty in quadrature to the photometry measurements for each filter before conducting the SED fitting. 
\begin{table}[htbp]
    \caption{\centering Parameters used for SED fitting with \textsc{bagpipes}}
    \label{tab:SED-fitting}
    \footnotesize
    \centering
    \setlength{\tabcolsep}{4pt} 
    
    \begin{tabularx}{\columnwidth}{XcX}
        \hline\hline
        Parameter & Fixed / Prior & Range\\
        \hline\\[-0.87em]
        Redshift ($z$) & Fixed & Spectroscopic redshift\\                 
        \hline\\[-0.87em]
        Metallicity ($Z$) & Logarithmic & $[0.05;3]\,\mathrm{Z_{\odot}}$ \\
        Ionisation parameter ($\log\mathcal{U}$) & Uniform & $[-4;-1]$\\
        Birth clouds age ($t_{bc}$) & Fixed & $\SI{10}{Myr}$\\
        \hline\\[-0.87em]
        \multicolumn{3}{c}{\textbf{Dust}} \\
        Extinction curve model & Fixed &\cite{Calzetti:2000}\\
        Dust emission model & Fixed &\cite{Draine:2007}\\
        V-band extinction ($A_V$) & Uniform & $[0;6]$ \\
        \hline\\[-0.87em]
        \multicolumn{3}{c}{\textbf{Continuity SFH}} \\
        Initial mass function & Fixed & \cite{Kroupa:2001} \\
        Stellar population & Fixed & \cite{Bruzual:2003} \\
        Age bins & Fixed & $(0,0.03,0.10,0.25,0.6,$ $1.4,3.2,7.5,10^{(a)})\,\mathrm{Gyr}$\\
        $\log\biggl(\frac{\mathrm{{SFR(t)}}}{\mathrm{SFR(t+\Delta t)}}\biggr)$ & Student-t & \makecell[l]{$\nu=2,\sigma=0.3$\\Physical $\mu^{(b)}$}\\
        $\log{(\text{Stellar mass}\,\mathrm{[M_\odot]})}$ & Uniform & $[6;15]$\\
        \hline
    \end{tabularx}
    \tablefoot{$^{(a)}$The last age bin ends at redshift $z=20$ for each galaxy.\\
    $^{(b)}$Prior parameters: $\nu$ is the wings weight, $\sigma$ is the scale and $\mu$ is the distribution location. The latter is set to follow the \citet{Madau:2014} SFR across cosmic time.}
\end{table}
The parameters employed in the SED fitting are summarised in \tabref{tab:SED-fitting}. We adopted a non-parametric continuity star formation history (SFH) to model complex SFHs. We set the scale parameter of the continuity SFH to $\sigma = 0.3$, as in \citet{Leja:2019}. We also set the prior on the change of the SFR across adjacent bins to follow the \citet{Madau:2014} cosmic star formation rate density.

Among all available bands, we excluded the filters contaminated by \ptre emission from the SED fitting for each galaxy. In particular, we excluded $\mathrm{F_C}$, $\mathrm{F_R}$, and $\mathrm{F_L}$. We also excluded the F444W filter for all galaxies and the F356W one for the galaxies at redshift $z<0.23$. We chose to exclude these bands to ensure a more straightforward comparison between the physical quantities derived from the SED fitting and those obtained directly from the \ptre emission. This approach enables us to compare the physical parameters inferred from SED fitting with those based on PAH luminosity in an independent and unbiased way, similarly to \citet{Benotto:2026a}.

\subsection{Colour--mass diagram}\label{sec:color-mass-division}
From the SED fitting, we also estimated the rest-frame $UBVJ$ AB-magnitudes of the galaxies, which we used to build the rest-frame $U-B$ colour (colour-- mass diagram) as a function of stellar mass and the UVJ diagram.
In the colour--mass diagram, to determine the separation lines of the green valley region, we first divide our sample into three mass bins: $[8.9, 9.4)$, $[9.4, 10.1)$, and $[10.1, 11.5]$. This division ensures that we have approximately the same number of galaxies in each bin. We then performed a fit using three Gaussian functions \citep[like in e.g.][]{Vulcani:2015} for each mass bin sample, constraining the centres of the three Gaussians to shift according to the same linear relationship as a function of mass. Finally, we defined the green sequence as the area that lies $2\sigma$ above the centre of the blue cloud in the $U-B$ colour space and $2\sigma$ below the red sequence. In particular, galaxies are assigned to the red sequence if:
\begin{equation}
    U-B>0.142 \log M_\star-0.293
\end{equation}
and to the blue cloud if
\begin{equation}
    U-B<0.142 \log M_\star-0.522
\end{equation}
where $\log M_\star$ is the base-10 logarithm of the stellar mass, as retrieved from SED fitting.

\subsection{UVJ diagram}\label{sec:transition-division}
We also analyse the UVJ diagram \citep{Williams:2009, Whitaker:2011}. Rather than adopting a sharp division between star-forming and quiescent galaxies, we define a transition region around the standard UVJ separation line. Galaxies close to this boundary have been shown to have intermediate properties and to broadly correspond to galaxies in the green valley of the colour--mass diagram \citep[e.g.,][]{Patel:2011, Arnouts:2013, Fang:2018, Gu:2018}. Our approach is similar to that adopted by \citet{McNab:2021} in the NUV-V V-J diagram and by \citet{Mao:2022}. Specifically, we defined the transition zone as the region extending within $\pm$0.1 magnitudes on either side of the standard diagonal UVJ separation line. So that:
\begin{equation}
      0.88(V-J) +0.59 < (U-V)_{\mathrm{ trans}} < 0.88(V-J) + 0.79
\end{equation}

\section{Results}\label{sec:results}
\begin{figure}
    \centering
    \includegraphics[width=\linewidth]{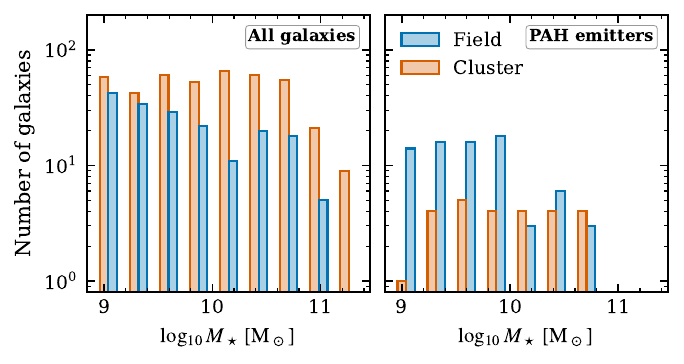}
    \caption{Number of galaxies per mass bin. The left panel shows the distribution for the entire galaxy sample, while the right panel shows the mass distribution of the PAH emitters only.}
    \label{fig:mass-distribution}
\end{figure}

The resulting galaxy sample, selected to have the required photometric bands and secure redshifts in the range $0.195<z<0.415$, and obtained after excluding known AGNs and imposing the mass completeness threshold, comprises 606 galaxies; of these, 425 are in clusters, and 181 are field galaxies. The left panel of \figref{fig:mass-distribution} displays the mass distribution of the sample. 
\begin{figure*}[hbt]
    \centering
    \includegraphics[width=0.9\linewidth]{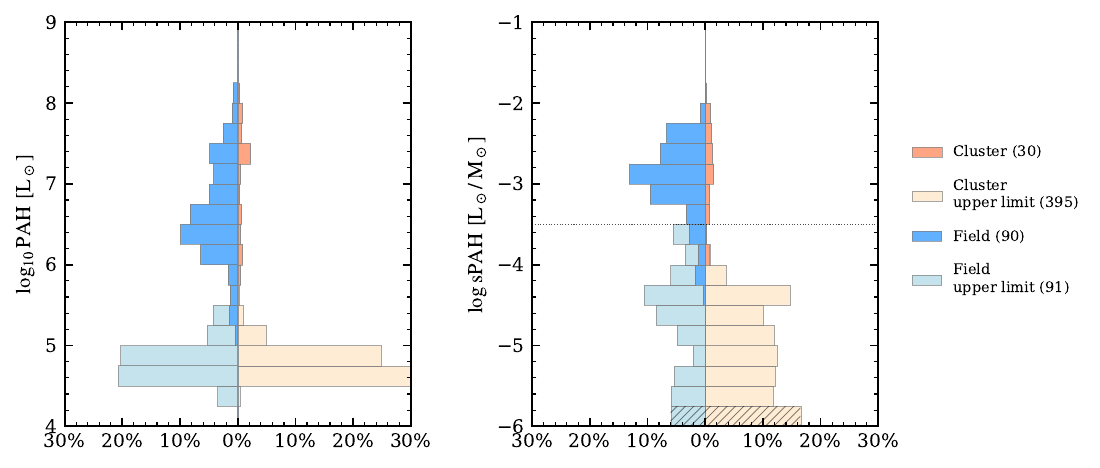}
    \caption{Distribution of PAH (left panel) and of the sPAH (right panel) luminosities. Both distributions are normalised to the total number of galaxies in both the field (blue) and the cluster (orange). The lighter colours represent the galaxies for which the PAH is not detected. The hatched bin includes all the galaxies with $\mathrm{sPAH}< 10^{-5.75}\,\mathrm{L_\odot/M_\odot}$. The horizontal dashed line represents the emitting galaxy threshold considered in this work.}
    \label{fig:PAH}
\end{figure*}
Figure~\ref{fig:PAH} shows the total \ptre luminosity distribution for the galaxies in our sample. The figure also displays the distribution of the specific PAH luminosity (sPAH), defined as the ratio of \ptre luminosity to the host galaxy's stellar mass, as derived from SED fitting. In the figure, we also display the upper limit values for galaxies where no PAH emission has been detected.

From now on, we consider a galaxy to be PAH emitting whenever its sPAH luminosity exceeds $\mathrm{sPAH}> 10^{-3.5}\,\mathrm{L_\odot/M_\odot}$.
The choice to use sPAH rather than total PAH luminosity for the cut is motivated by the need to isolate the intrinsic physical state of the interstellar medium from the galaxy's overall scale. Using integrated PAH luminosity would result in a sample dominated by massive galaxies, as lower-mass systems would require extreme dust-to-mass ratios to reach the same detection threshold.
The chosen cut is visible in \figref{fig:PAH} as a dashed line, and ensures that every galaxy upper limit has a sPAH below the threshold. Sixteen galaxies have been identified with detectable PAH regions, yet their global sPAH levels fall below the established threshold. In most of these cases, PAH emissions are observed in visible dust lanes or are concentrated in the central regions of the galaxies, which may suggest the presence of a younger star population in the galactic bulge or the existence of unidentified AGNs.

The stellar mass distribution of the \ptre emitting galaxies is shown in the right panel of \figref{fig:mass-distribution}, which displays the mass distribution of both the entire galaxy sample and the PAH emitters. A steep drop in PAH emitters is observed for $\log M_\star/M_\odot < 9.5$ in clusters, while a factor $\sim4$ drop in PAH emitters for field galaxies is observed at $\log M_\star/M_\odot >10$.

The average sPAH in our sample for the emitting galaxies is $\bigl<\mathrm{\log_{10}sPAH\ [L_\odot/ M_\odot]}\bigr> = -2.8$ and the values are compatible with a constant when computed after dividing our sample into mass bins, even though the scatter in each bin is of $\sim\SI{0.4}{dex}$. This value consistently holds for both cluster and field galaxies over more than 2~dex of stellar masses. Our final sample of emitting galaxies comprises $(42\pm4)\%$ of the field galaxies and $(6\pm1)\%$ of the cluster galaxies: this large difference between the two global fractions already points to an environmental effect, though it simply reflects the different mix of star-forming and quiescent galaxies in the two environments, rather than a direct effect on the PAH molecules. The fraction of emitting galaxies for star-forming and quiescent galaxies is analysed in \secref{sec:fraction-quiescent}.

\subsection{Rest-frame U-B vs stellar mass}\label{sec:cmd}
\begin{figure}[htb]
    \centering
        \includegraphics[width=0.8\linewidth]{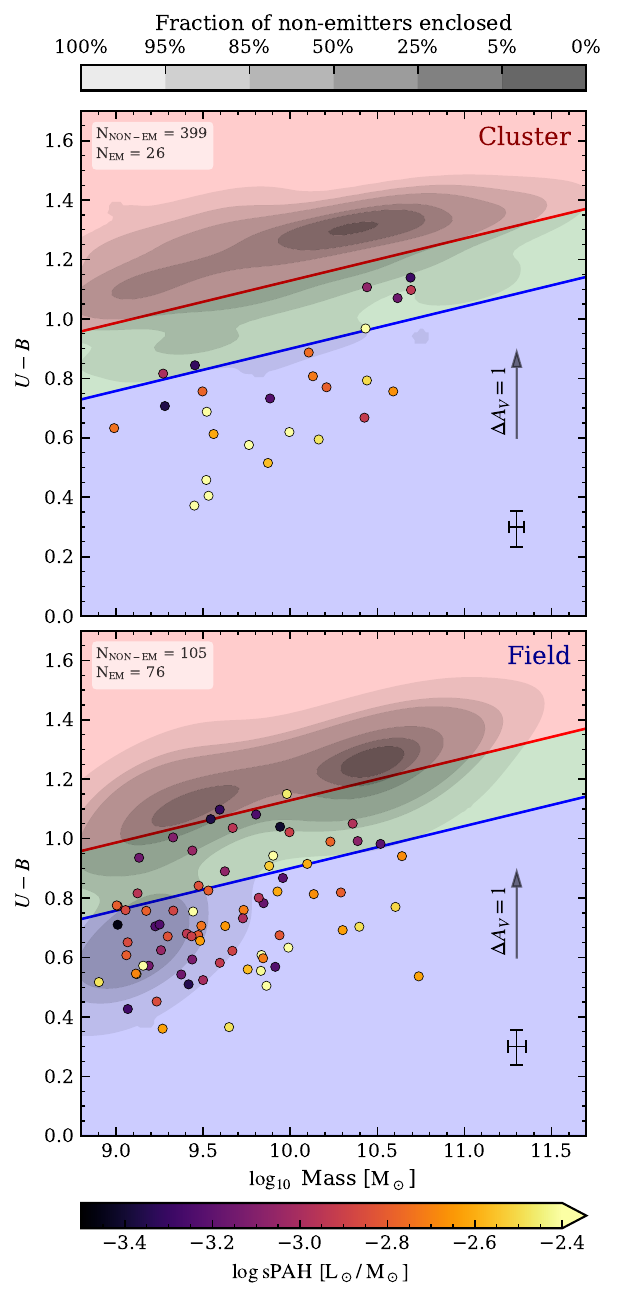}
    \caption{$U-B$ vs stellar mass diagram from SED fitting. The upper and lower panels display the cluster and field galaxies, respectively. The black contours represent the distribution of non-emitting galaxies, while each coloured circle depicts an emitting galaxy. The distribution is uncorrected for completeness. The total number of emitting ($\mathrm{N_{EM}}$) and non-emitting ($\mathrm{N_{NON-EM}}$) galaxies for each panel is visible in the upper left corner. The median error bars for the emitting galaxies are shown in the lower-left corners. The background colours indicate the blue, green and red galaxy sequence, while the arrow displays the shift in the diagram when the dust extinction in the $V$ band increases by 1 magnitude.}
    \label{fig:cmd}
\end{figure}
To analyse the PAH properties of the galaxies as a function of their physical properties, we first consider the $U-B$ vs stellar mass diagram for cluster and field galaxies separately, which is displayed in \figref{fig:cmd}. The cluster galaxies colour--mass diagram reveals a lack of blue faint ($\log M_\star<9.5$) galaxies, which are instead abundant in the field. This difference may reflect the impact of the cluster environment, as low-mass galaxies are more easily and rapidly deprived of the gas required to sustain star formation \citep{Mao:2022, Oxland:2024}.
Among field galaxies, at similarly low stellar masses, we observe a large scatter in sPAH, with some galaxies showing strong PAH emission and many others having undetectable sPAH. This diversity may be related to the broad range of metallicities spanned by low-mass galaxies, since low-metallicity systems are known to exhibit a PAH deficit \citep{Whitcomb:2024, Lai:2025}. In the green valley, $(43\pm6)\%$ of  field galaxies emit PAH, compared to only $(7\pm2)\%$ in  clusters. Similarly, no red cluster galaxies are detected as PAH emitters, whereas $(3\pm1)\%$ of red field galaxies show detectable PAH emission. The contrast between the cluster and field populations is therefore particularly striking in the green valley, and persists into the red sequence. However, these trends should be interpreted with caution, as both the green valley and red sequence may include dusty star-forming galaxies whose red colours do not necessarily indicate a genuinely quiescent stellar population \citep{Brammer:2009, Suess:2021}.

\subsection{UVJ diagram}\label{sec:uvj}
\begin{figure}[htb]
    \centering
    \includegraphics[width=0.8\linewidth]{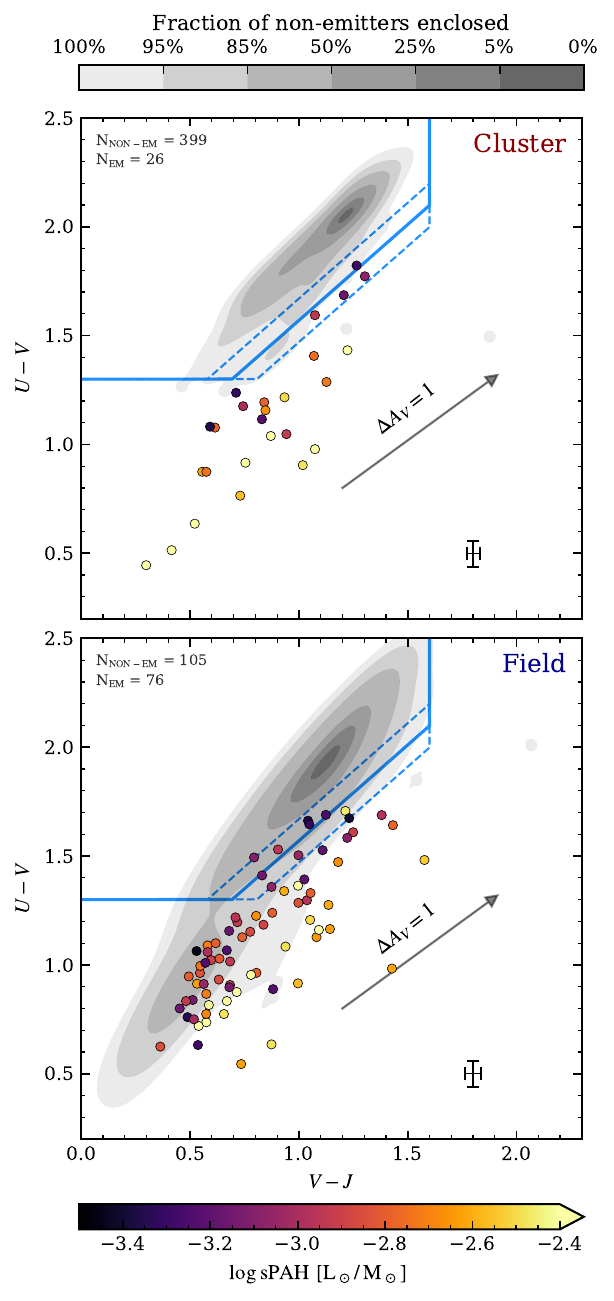}
    \caption{UVJ diagram for cluster (top panel) and field galaxies (bottom panel). The solid light-blue line delimits the star-forming region from the quiescent one as defined in \citet{Whitaker:2011}. The dashed line delimits the transition region as defined in the text. All the other quantities are displayed as in \figref{fig:cmd}.}
    \label{fig:uvj}
\end{figure}

To disentangle the effects of galaxy age and dust extinction, we also analyse the UVJ diagram \citep{Williams:2009, Whitaker:2011}, shown in \figref{fig:uvj}. 
Every \ptre emitter lies in the star-forming region or in the transition zone. Conversely, we find no \ptre emission from two galaxies located in the upper-right region of the UVJ diagram. This region is generally associated with highly dust-obscured star-forming systems. However, the SED fitting indicates very low metallicities for these galaxies, with  $Z=\SI{0.06(2)}{Z_\odot}$, which may suppress PAH emission and explain its non-detection \citep{Whitcomb:2024}.

The bluest part of the UVJ star-forming region ($V-J < 0.7$) is primarily populated by low-mass galaxies, with an average stellar mass of $\langle\log M_\star/M_\odot\rangle = 9.3\pm0.3$. This region is sparsely populated in the clusters,  with only nine detected galaxies compared to 61 for field galaxies. Approximately $(61 \pm 6)\%$ of the field galaxies in this region are \ptre emitters.

Finally, a clear environmental difference is apparent in the transition zone. In the field, this region contains a substantial population of $\mathrm{PAH_{3.3}\text{-emitting}}$ galaxies, whereas the corresponding cluster population is dominated by non-emitters. The result found here reflects what we found in the green valley in \figref{fig:cmd}, and proves that those results persist even when the dust effects are mitigated. We explore the observed difference in detail and its possible implications in \secref{sec:transition}.

\subsection{Fractions of PAH emitters}\label{sec:fractions}
\begin{figure*}[htbp]
    \centering
    \renewcommand{\arraystretch}{1}
    \setlength{\tabcolsep}{4pt}
    \begin{minipage}[b]{0.48\linewidth}
      \centering
      \colorbox{black!9}{%
        \begin{overpic}[width=0.98\linewidth]{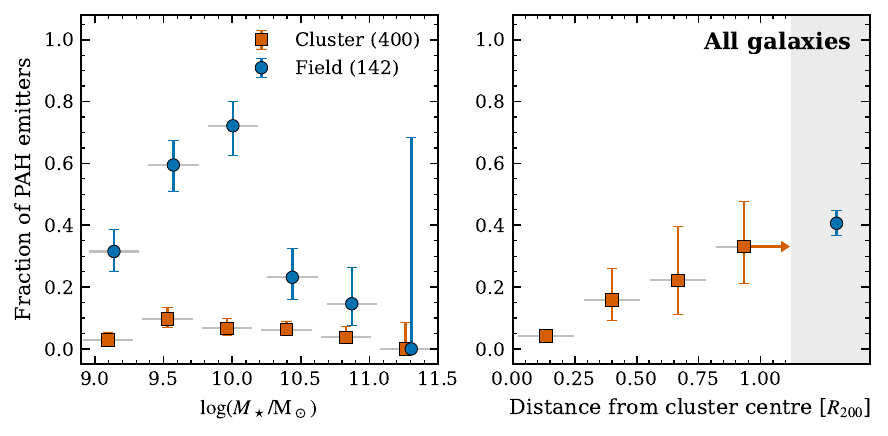}
          \put(7,3){\makebox[0pt][r]{\color{black}\textbf{(a)}}}
        \end{overpic}%
      }
      \subfiglabel{fig:a}
    \end{minipage}
    \hfill
    \begin{minipage}[b]{0.48\linewidth}
      \centering
      \colorbox{black!9}{%
        \begin{overpic}[width=0.98\linewidth]{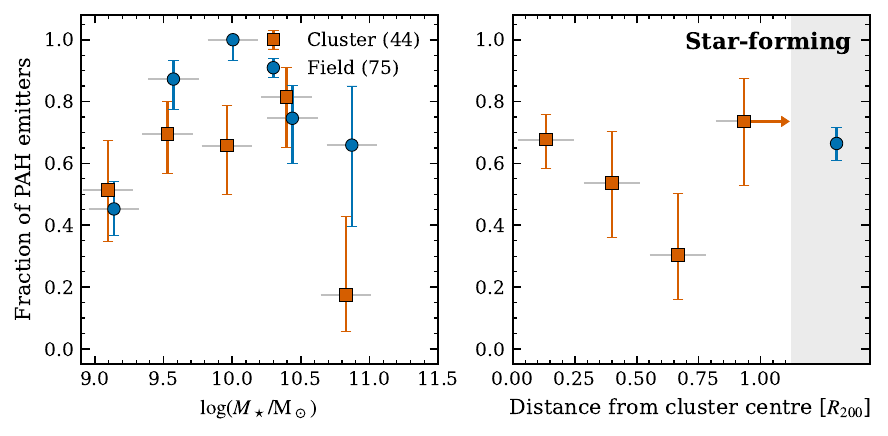}
          \put(7,3){\makebox[0pt][r]{\color{black}\textbf{(b)}}}
        \end{overpic}%
      }
      \subfiglabel{fig:b}
    \end{minipage}

    \vspace{0.6em}

    \begin{minipage}[b]{0.48\linewidth}
      \centering
      \colorbox{black!9}{%
        \begin{overpic}[width=0.98\linewidth]{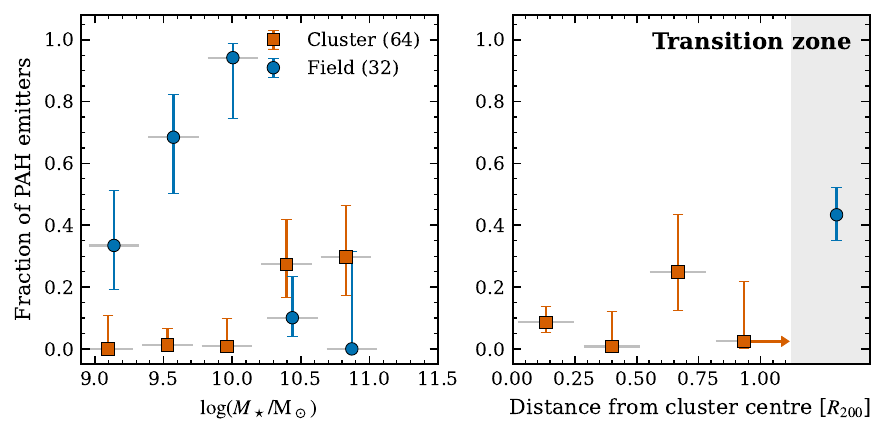}
          \put(7,3){\makebox[0pt][r]{\color{black}\textbf{(c)}}}
        \end{overpic}%
      }
      \subfiglabel{fig:c}
    \end{minipage}
    \hfill
    \begin{minipage}[b]{0.48\linewidth}
      \centering
      \colorbox{black!9}{%
        \begin{overpic}[width=0.98\linewidth]{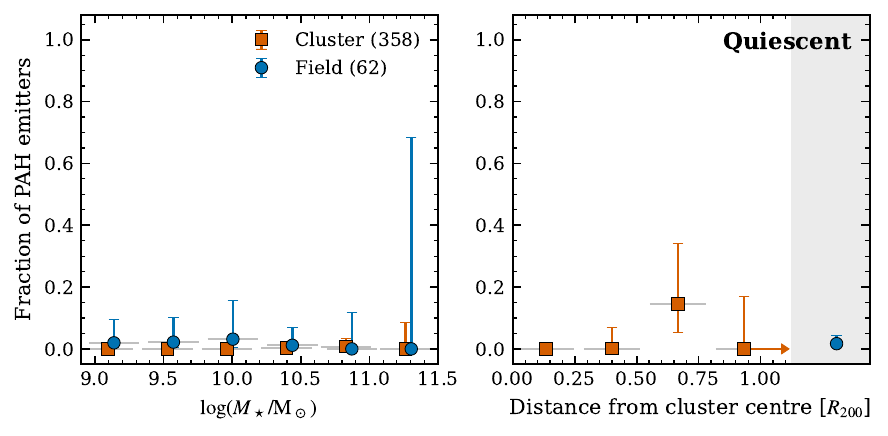}
          \put(7,3){\makebox[0pt][r]{\color{black}\textbf{(d)}}}
        \end{overpic}%
      }
      \subfiglabel{fig:d}
    \end{minipage}
    \caption{Fraction of PAH emitters as a function of stellar mass and cluster-centric distance. The distribution of PAH emitter fractions is shown for the entire galaxy sample (a), for the UVJ star-forming galaxies (b), for the galaxies located in the UVJ transition zone (c), and the quiescent UVJ galaxies (d). In each graph, the fraction is displayed for the clusters in orange and the field in blue. The grey zone in the cluster-centric panels separates the field's overall fraction from the cluster fractions. In the label, the effective number of galaxies for the field and cluster in that zone is shown in the legend.}
    \label{fig:fractions}
\end{figure*}

To better understand whether the \ptre properties of cluster and field galaxies differ from each other, we analyse the fraction of \ptre emitting galaxies as a function of mass and cluster-centric distance (\figref{fig:fractions}). Since the three clusters are not virialised, we consider multiple centres for each cluster, and we define the cluster-centric distance as the minimum among the distances from each centre. To compare the clusters together, we also normalised the distances to the cluster virial radius ($R_{200}$). The centres used and $R_{200}$ are summarised in Appendix~\ref{app:centers}.

For each galaxy, we estimate the probability of belonging to the star-forming, quiescent, or transition region of the UVJ diagram through a Monte Carlo analysis. Assuming Gaussian asymmetric uncertainties on the rest-frame $U-V$ and $V-J$ colours, we generate $N=1000$ realisations of each galaxy's position in the UVJ plane by randomly perturbing its colours within their errors, and for each realisation we classify the galaxy as star-forming, quiescent, or in the transition zone according to its position relative to the separation line defined earlier. The fractions shown in \figref{fig:fractions} are computed by weighting each galaxy both by its completeness correction and by these region-membership probabilities; the associated uncertainties on the fraction further include the binomial contribution from the sample size. 

\subsubsection{Quiescent, star-forming and overall fraction}\label{sec:fraction-quiescent}
Figure \ref{fig:fractions} shows the fraction of PAH emitters in each of these groups: all galaxies, star-forming, quiescent and transition zone. In particular, Figure \ref{fig:a} displays the fraction for all galaxies. We note that, at any given stellar mass, the fraction of \ptre emitting galaxies in the entire galaxy sample is higher in the field than in clusters, as already discussed at the beginning of \secref{sec:results} and in \figref{fig:PAH}.
For both cluster and field galaxies, the fraction of emitters as a function of stellar mass peaks at around $10^{9.5}$--$10^{10}\, \mathrm{M_\odot}$. The decline at higher masses can be largely attributed to the increasing proportion of quiescent galaxies in this mass range. Indeed, when considering star-forming galaxies alone (\figref{fig:b}), the decrease is much less pronounced. 

In the lowest mass bin of \figref{fig:a}, the reduced fraction of \ptre emitters reflects two distinct effects in the field and in clusters. Restricting to star-forming galaxies alone, the decline in the field is the same, indicating that at low masses galaxies have intrinsically lower \ptre emission on average. This may be related to the mass--metallicity relation \citep{Tremonti:2004, Maiolino:2019, Andrade:2024} and to the associated suppression of PAH features in low-metallicity systems \citep{Whitcomb:2024}, as already found in \secref{sec:cmd}. In clusters, by contrast, the low-mass bin is dominated by quiescent galaxies, as star-forming galaxies are essentially absent below $\log M_\star/M_\odot < 9.25$: only two cluster galaxies in our sample have a probability greater than 50\%  of being star-forming in this mass range. To complete the analysis of this lowest mass bin, the observed depletion of PAH emitting galaxies is not simply a completeness effect, as the reduction remains consistent even when the mass limit is varied by $\pm\SI{0.3}{dex}$.

Considering the dependence on cluster-centric distance in \figref{fig:a}, the fraction of \ptre emitters among all cluster members increases steadily towards the cluster outskirts, reaching a value $1\sigma$ compatible with that of the field at $R > 0.8\,R_{200}$. When considering star-forming, quiescent and transition-zone galaxies separately, however, no clear trend is visible as a function of cluster-centric distance. For this reason, the trend observed for all galaxies together simply reflects a higher fraction of quiescent galaxies towards the cluster centre \citep[e.g.][]{Vulcani:2013, Pintos-Castro:2019}. 

Quiescent galaxies (\figref{fig:d}) are almost always non-emitters, apart from some residual probability of galaxies in the transition zone of the cluster plot to be in the quiescent zone. Given the ability of PAH to trace SFR, this result is completely expected.

\subsubsection{The transition zone}\label{sec:transition}
\begin{figure}
    \centering
    \includegraphics[width=\linewidth]{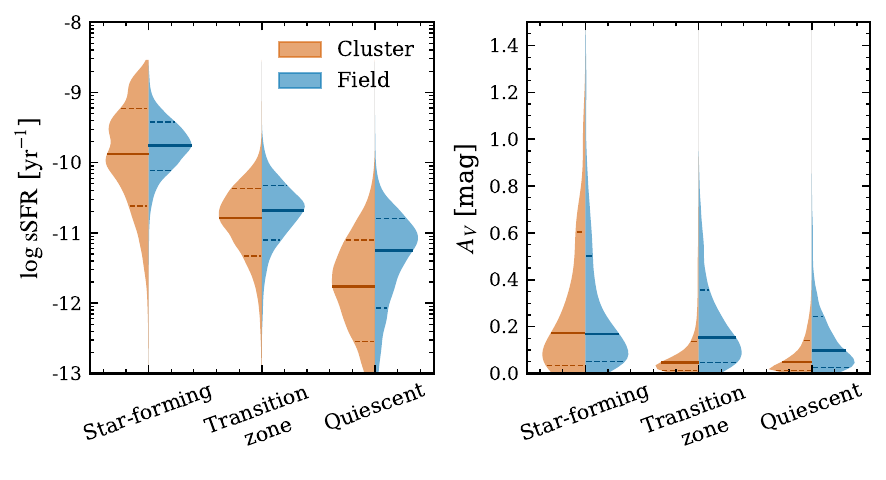}
    \caption{Distribution of sSFR and $A_V$ for cluster and field galaxies in the three UVJ regions defined in the text. The solid line represents the median value of the distribution, while the dashed lines indicate the $\nth{16}$ and $\nth{84}$ percentiles.}
    \label{fig:violin}
\end{figure}
The clearest difference between cluster and field galaxies emerges in the transition zone (\figref{fig:c}), where the behaviour depends strongly on stellar mass. Below $\log M_\star/\mathrm{M_\odot} < 10.2$,  $63_{-12}^{+10}\%$ of field galaxies are \ptre emitters, compared to only $1_{-1}^{+3}\%$ of cluster galaxies, which represent a $5\sigma$ significance result. The fraction of transition zone emitters in clusters remains significantly below that in the field out to $R = R_{200}$. At fixed stellar mass, transition-zone galaxies in the field show an emitter fraction comparable to that of the overall star-forming field population up to $\log M_\star/\mathrm{M_\odot} < 10.2$ (\figref{fig:b}).

Since \ptre emission is commonly used as a tracer of recent, dusty star formation, the observed deficit might naively be attributed to lower star-formation activity in cluster galaxies. However, results from the SED fitting show that the sSFR --- weighted for completeness and for the probability of membership in each zone --- steadily decreases from the star-forming region to the quiescent region through the transition zone (\figref{fig:violin}), consistent with previous studies \citep[e.g.,][]{Patel:2011, Fang:2018}. In both star-forming, quiescent, and transition zones, the distributions for cluster and field galaxies differ but remain broadly compatible, with the transition zone showing the closest agreement between the two environments ($\log\mathrm{sSFR}\ \mathrm{[yr^{-1}]}=-10.7_{-0.4}^{+0.4}$ for field galaxies versus $-10.8_{-0.5}^{+0.4}$ for cluster galaxies.
To evaluate whether the similarities or differences observed in the field-vs-cluster distributions are statistically significant, we perform a weighted Kolmogorov-Smirnov (KS) test \citep{Kolmogorov:1933, Smirnov:1948}. The implemented algorithm accounts for both completeness and membership probabilities as weights. The $D_{\mathrm{KS}}$ statistic represents the maximum absolute vertical distance between the two weighted empirical cumulative distribution functions. In the case of $\log\mathrm{sSFR}$, we obtain compatibility in the transition zone with $D_{\mathrm{KS}} = 0.13$ and a p-value of $p = 0.15$ for the same-distribution hypothesis.
\begin{figure*}     
    \centering     
    \includegraphics[width=0.85\linewidth]{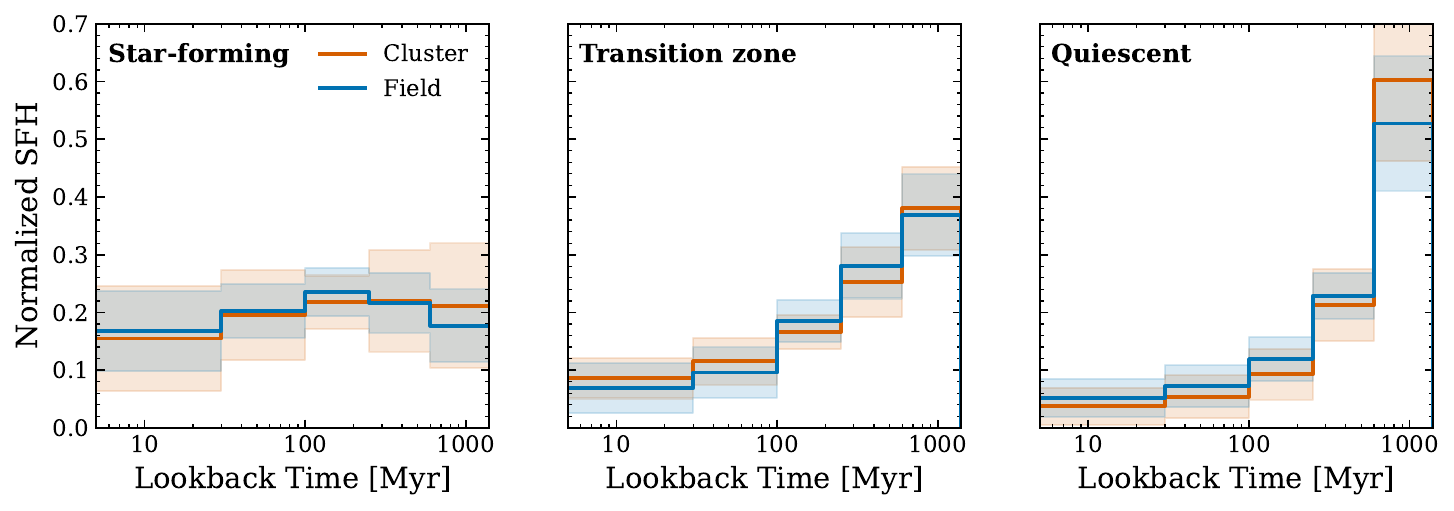}     
    \caption{Average normalised SFH for the field and cluster galaxies, in the UVJ star-forming, transition, and quiescent zones. Each SFH is normalised by the sum of the SFH in bins younger than 1~Gyr to show the average shape of the SFH, rather than the effective SFR at any time. The colored zones represent the standard deviation of the normalised SFR in that bin.}     
    \label{fig:sfh} 
\end{figure*}

Moreover, the average SFH of transition-zone galaxies is also indistinguishable between field and clusters. In \figref{fig:sfh} we show the normalised SFH for all galaxies with masses below $10^{10.2}\,\mathrm{M_\odot}$, which are the masses where the fraction of PAH emitters between clusters and fields is more different. In the transition zone, both SFHs show a gradual decline while retaining some star formation over the last \SI{100}{Myr}.

The $A_{V}$ distributions, visible in the right panel of \figref{fig:violin}, reveal an even more interesting pattern. Among star-forming galaxies, the $A_{V}$ distributions of cluster and field galaxies are nearly indistinguishable ($D_{\mathrm{KS}} = 0.06$, $p = 0.69$), while the two environments diverge sharply in the transition zone ($D_{\mathrm{KS}} = 0.44$, $p < 10^{-3}$). Field galaxies in the transition zone retain an $A_{V}$ distribution similar to that of star-forming galaxies ($D_{\mathrm{KS}} = 0.09$, $p = 0.32$), whereas cluster galaxies show a marked drop, with an $A_V$ in the transition zone already comparable to the distribution seen in the cluster quiescent region ($D_{\mathrm{KS}} = 0.04$, $p = 0.31$).

Taken together, these results demonstrate that while stellar populations evolve smoothly toward older ages as galaxies move from the star-forming to the quiescent zone in both environments, their ISM follow distinct evolutionary paths. In the field, the dust content remains comparable between the star-forming and transition regimes, decreasing only modestly in the quiescent phase, whereas cluster galaxies encounter an abrupt environmental ISM change when entering the transition zone. Crucially, this divergence implies that the lower fraction of \ptre emitters in cluster transition galaxies (\figref{fig:c}) cannot be attributed merely to reduced star-formation activity, but rather points to a direct suppression of PAH emission at a fixed level of star formation likely driven by altered ISM conditions or accelerated dust processing \citep{Cortese:2010, Cortese:2012}. To further explore how these interconnected physical properties influence PAH emission, we present the dependence of the emitting fraction as a function of both dust extinction and metallicity in Appendix~\ref{app:AvZ-dep}.

\section{Discussion}\label{sec:discussion}
This work investigates the integrated \ptre emission of field and cluster galaxies. We find a cluster $\mathrm{PAH_{3.3}}$-emitter fraction of \SI{6(1)}{\%}, which is substantially lower than the $10$--$25\%$ blue galaxy fraction typically expected at comparable redshifts and masses \citep{Wetzel:2012, Haines:2013, hashimoto:2019}. While these metrics are not perfectly equivalent due to varying definitions of active galaxies, this discrepancy highlights a remarkably low incidence of \ptre emission in our cluster sample. Part of this difference might be related to the incomplete radial coverage of the clusters: A2744 is sampled only out to $0.25\,R_{200}$, while the coverage of MACS0416 at  $R>R_{200}$ is also limited. Since the fraction of star-forming galaxies is known to decrease towards the cluster core \citep[e.g.,][]{Balogh:1999, Cohen:2014}, our observations preferentially probe the regions with the lowest fraction of star-forming galaxies, where a lower incidence of \ptre emission is therefore also expected. The limited radial coverage may consequently bias the integrated cluster \ptre emitting fraction towards lower values.

Our field fraction of PAH emitters, $\SI{42(4)}{\%}$, is instead in good agreement with the $\sim47\%$ (88/187) detection fraction reported by \citet{Lyu:2025} for a similarly-selected, similarly-distant ($z\sim0.2-0.5$) field galaxy sample using JWST/NIRCam grism spectroscopy. This consistency, obtained with independent selection functions and PAH-detection methods (integrated medium-band photometry versus spectroscopic $\SI{3.3}{\micro\meter}$ line detection), supports the reliability of our measurements and reinforces that the deficit we observe in clusters is a real environmental effect. \citet{Lyu:2025} additionally find that the \ptre versus IR luminosity ratio drops sharply below $12+\log(\mathrm{O/H})\sim8.4-8.5$, which offers a natural explanation for the decline in our field PAH fraction towards lower stellar masses (Sect.~\ref{sec:results}): low-mass, metal-poor field galaxies are expected to have less PAH emission at constant star-formation rate, independently of any environmental effect \citep{Whitcomb:2024}.

From our clustercentric analysis, in which the PAH-emitting fraction rises from $(5\pm1)\%$ at $R<0.25\,R_{200}$ to $(33\pm15)\%$ at $R=0.8 R_{200}$, also emerges a result consistent with the results of \citet{Cheng:2025}, who identified a sample of 3.3\,$\mu$m PAH-bright galaxies in A2744 using JWST medium-band imaging. They find that PAH emitters are preferentially located in regions of relatively low mass surface density within the cluster, with SFRs consistent with the field main-sequence at $z=0.3$, and interpret this population as galaxies that are falling into the cluster and have not yet been quenched. Our radial trend independently supports this scenario: PAH emission survives preferentially in the cluster outskirts, where galaxies have spent, on average, less time subject to cluster environmental processes.

The low PAH fraction we measure is especially pronounced in the UVJ transition zone, where cluster and field galaxies have statistically indistinguishable star-formation histories over the past Gyr (see \secref{sec:transition}), yet only $1_{-1}^{+3}\%$ of cluster transition galaxies show PAH emission, against $63_{-12}^{+10}\%$ in the field. This evidence indicates that the deficit originates in the state of the ISM itself, rather than in any difference in the galaxies' SFHs. The declining SFH of these transition galaxies resembles that of mid-old infallers, which experienced a star formation peak in the past and now have little star formation remaining; their suppression is consistent with environmental processes acting within the cluster, such as tidal interactions \citep{Moore:1996}, starvation or strangulation \citep{Larson:1980}, and ram-pressure stripping (RPS; \citealt{Gunn:1972}).

Among these mechanisms, RPS provides a compelling explanation for the abrupt PAH deficit observed. \citet{Benotto:2026a} showed that PAH molecules can be actively stripped by RPS, a process capable of rapidly and almost completely clearing PAHs out of a galaxy. In this scenario, as soon as the SFR begins to decline, few PAH molecules remain to trace it, which would explain why cluster transition galaxies show essentially no PAH emission despite still retaining sSFRs comparable to their field counterparts. This interpretation is further supported by the lower median $A_V$ we measure in cluster transition galaxies relative to the field (\secref{sec:transition}), indicating that these galaxies have already lost a substantial fraction of their dust content by the time they reach the transition zone.

Taken together, our results show that PAH emission and dust content are suppressed by the cluster environment much faster than any change in the underlying SFH. To confirm this scenario directly, a tracer of star formation independent of dust and PAH emission is required, such as UV continuum light; in our sample, these data are only available for a very small number of galaxies, given the limited HST/UVIS F225W and F275W coverage. The results obtained in this work have direct implications for the use of PAH emission as an SFR tracer: calibrations based solely on intrinsic galaxy properties, such as mass or SFH, are insufficient in cluster environments, where the suppression of PAH features can lead to an underestimation of the true SFR unless environment is explicitly taken into account.

\section{Summary and conclusions}\label{sec:conclusions}

We analysed the integrated \ptre emission of 606 galaxies (425 cluster members, 181 field galaxies) at $0.195 < z < 0.415$ using JWST/NIRCam medium-band photometry. Stellar masses, rest-frame UVJ colours, and SFRs were derived through SED fitting with a non-parametric star-formation history and used to compare the incidence and properties of PAH-emitting galaxies in clusters and in the field. Our main results are as follows.

\begin{itemize}
    \item PAH emitting galaxies, defined as those with $\mathrm{sPAH} >10^{-3.5}\,\mathrm{L_\odot/M_\odot}$, are the $\SI{42(4)}{\%}$ of field galaxies but only $\SI{6(1)}{\%}$ of cluster members, highlighting a strong environmental suppression of the \ptre feature emission.
    \item In the $U-B$ vs mass diagram, field PAH emitters span the blue, green, and (occasionally) red regions, whereas PAH-emitting cluster galaxies are confined almost exclusively to the blue region: only four galaxies lie substantially within the green valley, all with $M>10^{10.5}\, M_\odot$.
    \item Cluster galaxies with $M<10^{9.5}\, M_\odot$ are essentially absent from the blue cloud, while $44$ such galaxies are found in the field, pointing to an efficient quenching of low-mass star-forming galaxies in clusters.
    \item The PAH-emitting fraction depends on stellar mass in both environments, peaking at $(72\pm8)\%$ around $M=10^{10}\, M_\odot$ for field galaxies. For these galaxies, it declines toward lower masses, to $(32\pm7)\%$ at $M<10^{9.5}\, M_\odot$, plausibly reflecting the lower metallicities of these systems, and toward higher masses, to $(19\pm5)\%$ at $M\ge10^{10.5}\, M_\odot$, consistent with the growing contribution of quiescent galaxies at the high-mass end. Cluster galaxies follow the same qualitative trend but at a systematically lower normalisation, reflecting the overall suppression of PAH emission in the cluster environment.
    \item The PAH-emitting fraction in clusters increases with clustercentric distance from $(5\pm1)\%$ at $R<0.25 R_{200}$ to $(33\pm15)\%$ at $R\sim R_{200}$, which reflects the higher concentration of quiescent galaxies towards the cluster centre.
    \item We defined a transition zone in the UVJ diagram as a band around the star-forming/quiescent separation line of \citet{Williams:2009}, selecting galaxies close to this boundary on either side. Within this zone, the difference between the two environments is most pronounced for galaxies with $M < 10^{10.2}\, M_\odot$, where $63_{-12}^{+10}\%$ of field galaxies are PAH emitters versus $1_{-1}^{+3}\%$ of cluster galaxies. In this mass range, the star-formation histories of cluster and field galaxies are on average indistinguishable, implying that the deficit of PAH emission is driven by the state of the interstellar medium rather than by any difference in the stellar populations built up over the galaxies' histories.
\end{itemize}

Taken together, these results show that the effect of the cluster environment is visible in the PAH emission of galaxies independently of their stellar population properties, indicating that environmental mechanisms act directly on the interstellar medium. As a future perspective, a spatially resolved analysis of the PAH emission will help to establish where this emission is concentrated within galaxies in the two environments, and how this distribution depends on galaxy morphology.

\begin{acknowledgements}
We are grateful to the anonymous referee for their insightful comments which significantly improved this work. The authors would like to thank L. Bisigello for the stimulating discussions and valuable insights. This research is based on observations made with the NASA/ESA Hubble Space Telescope obtained from the Space Telescope Science Institute, which is operated by the Association of Universities for Research in Astronomy, Inc., under NASA contract NAS 5–26555. This work is also based on observations made with the NASA/ESA/CSA James Webb Space Telescope. The raw data were obtained from the Mikulski Archive for Space Telescopes at the Space Telescope Science Institute, which is operated by the Association of Universities for Research in Astronomy, Inc., under NASA contract NAS 5-03127 for JWST. The processed data products were retrieved from the DAWN JWST Archive (DJA). DJA is an initiative of the Cosmic Dawn Center (DAWN), which is funded by the Danish National Research Foundation under grant DNRF140. B. Vulcani and P. Watson also acknowledge support from the INAF Mini Grant ``1.05.24.07.01 RSN1: Spatially-resolved Near-IR Emission of Intermediate-Redshift Jellyfish Galaxies'' (PI Watson). 

\end{acknowledgements}
\bibliography{aa}

\begin{appendix}
\nolinenumbers
\section{Filters used in this work}\label{app:filters}
We show in Table \ref{tab:filters} the filters used in this work, both for SED fitting and/or PAH extraction. We do not list the filters that do not cover any of the galaxies analysed in this work. Based on the galaxy redshift, different filters among those in the table are used to extract the PAH emission map. Moreover, some filters close to the \ptre emission at the galaxy redshift are removed from the SED fitting, as described in \secref{sec:pah-extraction}.   

\begin{table}[hb]
    \caption{\centering Photometric observations used in this analysis.}
    \label{tab:filters}
    \centering
    \renewcommand{\arraystretch}{1.3}
    \begin{tabular}{l p{2.37cm} p{2.2cm}}
    \hline\hline
         \textbf{Instrument} & \textbf{Filters} & \textbf{Use} \\
         \hline
         HST/WFC3/UVIS & F225W, F275W, F336W, F350LP, F606W, F814W, F850LP& SED fitting \\
         \hline
         HST/ACS & F435W, F475W, F606W, F625W, F775W, F814W, F850LP & SED fitting \\
         \hline
         HST/WFC3/IR & F105W, F110W, F125W, F140W, F160W & SED fitting \\
         \hline
         JWST/NIRCAM & F070W, F090W, F115W, F140M, F150W, F162M, F164N, F182M,  F187N, F200W, F210M, F250M, F277W, F300M, F335M, F356W, F360M, F410M, F430M, F460M, F480M & SED fitting \\
         \hline
         JWST/NIRCAM & F335M, F360M, F410M, F430M, F460M, F480M & PAH maps and SED fitting\tablefootmark{(a)} \\
         \hline\hline
    \end{tabular}
    \tablefoot{\centering
\tablefoottext{a}{Depending on the galaxy redshift, according to \tabref{tab:filters_pah}.}
}
\end{table}

\section{PAH flux calculation}\label{app:formula}
This section presents the derivation of the formula used to estimate the PAH flux ($F_\mathrm{PAH}$ in $\rm erg\,s^{-1}\,cm^{-2}$) from the flux density measured in the $\mathrm{F_C}$ filter, which encompasses the PAH feature.

Let $f_\nu$ be the source spectral flux density, $f_{\nu,\mathrm{cont}}$ the continuum flux density, and $T_\lambda$ the dimensionless transmission curve of the $\mathrm{F_C}$ filter. The measured flux density, $f_{\mathrm{F_C}}$, is given by
\begin{align}\label{eqn:1}
    f_{\mathrm{F_C}}=\frac{\int f_\nu\, T_\lambda\frac{\dd \lambda}{\lambda}}{\int T_\lambda \frac{\dd \lambda}{\lambda}}=f_{\nu,\mathrm{cont}} + \frac{\int f_{\nu,\mathrm{PAH}}\,T_\lambda\frac{\dd \lambda}{\lambda}}{\int T_\lambda \frac{\dd \lambda}{\lambda}},
\end{align}
where $f_{\nu,\mathrm{PAH}}$ is the spectral flux density associated with the PAH feature.

Assuming the PAH feature is sufficiently narrow compared to the filter bandwidth\footnote{PAH features are broader than emission lines but approximately five times narrower than the JWST medium-band filters used.}, its spectral flux density can be approximated by a Dirac delta function:
\begin{align}\label{eqn:2}
    f_{\nu,\mathrm{PAH}} = F_{\mathrm{PAH}}\frac{c}{\lambda_\mathrm{PAH}^2}\,\delta(\lambda-\lambda_\mathrm{PAH}),
\end{align}
where $F_{\mathrm{PAH}}$ is the PAH flux, $c$ is the speed of light, and $\lambda_\mathrm{PAH}$ is the central wavelength of the feature. Substituting Eq.~\ref{eqn:2} into Eq.~\ref{eqn:1} and solving for $F_{\mathrm{PAH}}$ yields
\begin{align}\label{eqn:3}
    F_\mathrm{PAH} = (f_{\mathrm{F_C}}-f_{\nu,\mathrm{cont}})\frac{c\int T_\lambda \frac{\dd \lambda}{\lambda}}{\lambda_\mathrm{PAH} T_{\lambda_\mathrm{PAH}}}.
\end{align}

Following \citet{Benotto:2026a}, the continuum flux density in the $\mathrm{F_C}$ filter is estimated by linearly interpolating the flux densities measured in the $\mathrm{F_L}$ and $\mathrm{F_R}$ filters, which bracket $\mathrm{F_C}$. Thus,
\begin{align}
    f_{\nu,\mathrm{cont}} = (1-k)f_{\mathrm{F_L}}+k f_{\mathrm{F_R}},
\end{align}
where $f_{\mathrm{F_L}}$ and $f_{\mathrm{F_R}}$ are the measured flux densities, in Jy, in the $\mathrm{F_L}$ and $\mathrm{F_R}$ filters, respectively, and
\begin{align}
    k=\frac{\lambda_{\mathrm{F_C}}-\lambda_{\mathrm{F_L}}}{\lambda_{\mathrm{F_R}}-\lambda_{\mathrm{F_L}}},
\end{align}
with $\lambda_{\mathrm{F_C}}$, $\lambda_{\mathrm{F_L}}$, and $\lambda_{\mathrm{F_R}}$ denoting the pivot wavelengths of the corresponding filters. The final expression to retrieve the PAH flux therefore is
\begin{align}\label{eqn:pah}
    F_\mathrm{PAH} = \bigl(f_{\mathrm{F_C}}-(1-k)f_{\mathrm{F_L}}-k f_{\mathrm{F_R}}\bigr)\frac{c\int T_\lambda \frac{\dd \lambda}{\lambda}}{\lambda_\mathrm{PAH} T_{\lambda_\mathrm{PAH}}}.
\end{align}

Equation \ref{eqn:pah} is the one used in this work; however, when the analysed features always fall far from the filter edges, the equation above could be simplified. Indeed, for a narrow ($w_{\mathrm{F_C}}\ll\lambda_{\mathrm{F_C}}$), box-shaped transmission curve of the $\mathrm{F_C}$ filter, where $w_{\mathrm{F_C}}$ is its rectangular width, the integral can be approximated as
\begin{align}
    \int T_\lambda \frac{\dd \lambda}{\lambda}\approx T\frac{w_{\mathrm{F_C}}}{\lambda_{\mathrm{F_C}}},
\end{align}
where $T$ is the constant transmission within the filter. Assuming the PAH feature lies within the filter, so that $T_{\lambda_\mathrm{PAH}}=T$ and $\lambda_\mathrm{PAH}\approx\lambda_{\mathrm{pivot},\mathrm{F_C}}$, Eq.~\ref{eqn:pah} simplifies to
\begin{align}\label{eqn:4}
    F_\mathrm{PAH} \approx \frac{c}{\lambda_{\mathrm{F_C}}^2} \,w_{\mathrm{F_C}}\,\biggl(f_{\mathrm{F_C}}-(1-k)f_{\mathrm{F_L}}-k f_{\mathrm{F_R}}\biggr),
\end{align}
which coincides with the formula adopted by, among others, \citet{Gullieuszik:2020} and \citet{Benotto:2026a}.

Using the approximated \eqref{eqn:4} instead of \eqref{eqn:pah} in this work would have had an impact on the PAH fluxes. We considered redshift ranges in \tabref{tab:filters_pah} that coincide with redshifts where $T_{\lambda_\mathrm{PAH}}$ is equal to half the maximum transmission of the PAH filter for each filter combination. This means that PAH fluxes would have been underestimated by up to a factor of two.

\section{Clusters centers and virial radius}\label{app:centers}
In \secref{sec:fractions} we compute the fraction of PAH emitting galaxies as a function of the cluster-centric distance.
However, the three clusters analysed (A2744, A370 and MACS0416) are not virialised, and considering a unique centre of the cluster would result in biased results. For this reason, we considered multiple centres for each cluster. Moreover, we also normalised the distances to the cluster `virial' radius ($R_{200}$) to compare the clusters together, even if they have different intrinsic sizes. The centres used and $R_{200}$ are summarised in \tabref{tab:centres}.
\begin{table}[htb]
    \centering
    \caption{\centering Virial radius and cluster centres, together with their reference.}
    \label{tab:centres}
    \begin{tabular}{llll}
    \hline\hline\\[-0.87em]
    Cluster & $R_{200}$ & RA center & DEC center\\
    \hline\\[-0.87em]
    A2744 & 2.5~Mpc &  00:14:22.17 & -30:24:18.32\\
    \multicolumn{2}{l}{\citep{2212.04381}}&  00:14:20.80 & -30:24:00.85\\
                    &&  00:14:13.16 & -30:22:35.65\\
                    &&  00:14:10.08 & -30:22:19.90\\
                    &&  00:14:18.83 & -30:21:24.84\\
    \hline\\[-0.87em]
    MACS0416 & 2.5~Mpc & 04:16:09.16 & -24:04:03.04 \\
    \multicolumn{2}{l}{\citep{10.1051/0004-6361/202244834}} & 04:16:07.69 & -24:04:38.80\\
    \hline\\[-0.87em]
    A370 & 1.8~Mpc & 02:32:52.72 & -1:34:18.92 \\
    \multicolumn{2}{l}{\citep{10.3847/1538-4357/ad684a}}& 02:39:53.113 & -1:34:56.12\\
    \hline
    \end{tabular}
\end{table}

\section{$A_V$ and metallicity dependences}\label{app:AvZ-dep}
\begin{figure}[ht]
    \centering
    \includegraphics[width=0.8\linewidth]{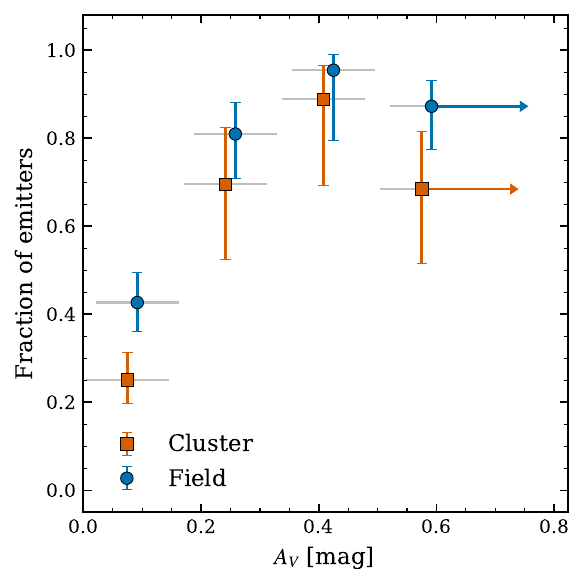}
    \includegraphics[width=0.8\linewidth]{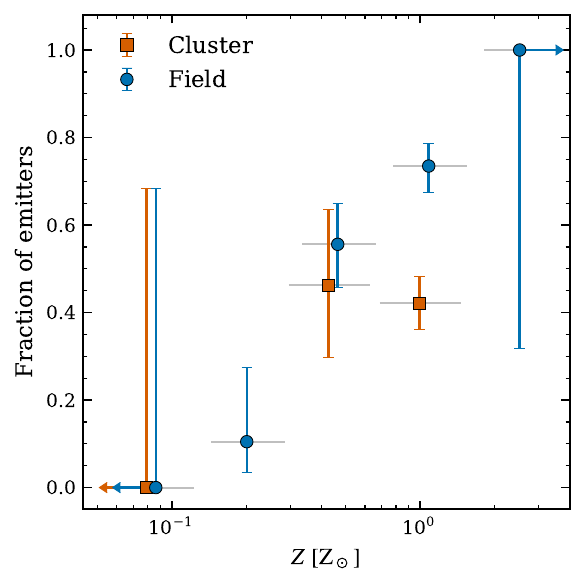}
    \caption{\ptre emitting galaxy fraction as a function of dust extinction (upper panel) and metallicity (lower panel). The graphs include the star-forming and transition galaxies together.}
    \label{fig:pah-av-z}
\end{figure}

In this work, we consider the fraction of emitters as a function of the galaxy masses and cluster-centric distance. However, we also analyse the galaxy emitting fraction as a function of dust extinction and metallicity for UVJ star-forming and transition zone galaxies (\figref{fig:pah-av-z}).
Using only the SED fitting results, we cannot overinterpret the trends due to the age-metallicity degeneracy. However, a general trend is noticeable: redder galaxies (those appearing redder because of dust extinction and/or metallicity) are more likely to be PAH emitters. In the dust extinction trend, the last bin contains only one galaxy from the field and one from the cluster that are not PAH emitters. These galaxies seem to deviate from the general trend; however, the statistics are limited, and these two galaxies are located in the upper-right region of the UVJ diagram, as discussed in Section \ref{sec:uvj}. No noticeable difference between cluster and field galaxies is observed with our statistics. The observed dependence of the emitting fraction as a function of dust attenuation qualitatively follows the observed trend in the UV-continuum slope vs. PAH mass fraction \citep{Mondal:2026} with a rapid growth at small $A_V$ and a steady, slowly decreasing function at high values of dust obscuration, supporting the reliability of our results.
\end{appendix}
\end{document}